# Comparison of the likelihood ratios of two diagnostic tests subject to a paired design: confidence intervals and sample size

José Antonio Roldán-Nofuentes[a], Saad Bouh Sidaty-Regad[b]

[a]Statistics (Biostatistics), School of Medicine, University of Granada, Spain

[b]Public Health and Epidemiology, School of Medicine, University of Nouakchott, Mauritania

Email: jaroldan@ugr.es, sidaty_saad@yahoo.com

**Abstract.** Positive and negative likelihood ratios are parameters which are used to assess and compare the effectiveness of binary diagnostic tests. Both parameters only depend on the sensitivity and specificity of the diagnostic test and are equivalent to a relative risk. This article studies the comparison of the likelihood ratios of two binary diagnostic tests subject to a paired design through confidence intervals. Six approximate confidence intervals are presented for the ratio of the likelihood ratios, and simulation experiments are carried out to study the coverage probabilities and the average lengths of the intervals considered, and some general rules of application are proposed. A method is also proposed to determine the sample size necessary to estimate the ratio between the likelihood ratios with a determined precision. The results were applied to two real examples.

# 1. Introduction

A diagnostic test is a medical test that is applied to an individual in order to determine the presence or absence of a disease. When the result of a diagnostic test is positive or negative, the diagnostic test is called a binary diagnostic test (*BDT*). A stress test for the diagnosis of coronary disease is an example of *BDT*. The effectiveness of a *BDT* is measured in terms of two fundamental parameters: sensitivity and specificity. The sensitivity (*Se*) is the probability of the *BDT* being positive when the individual has the disease, and the specificity (*Sp*) is the probability of the *BDT* being negative when the individual does not have it. The *Se* and the *Sp* of a *BDT* are estimated in relation to a gold standard (*GS*), which is a medical test which objectively determines whether or not an individual has the disease or not. An angiography for coronary disease is an example of *GS*. Other parameters that are used to assess the effectiveness of a *BDT* are the likelihood ratios (*LRs*) (Pepe, 2003; Zhou et al, 2011). When the *BDT* is positive, the likelihood ratio, called the positive likelihood ratio $\left(LR^+\right)$, is the ratio between the probability of correctly classifying an individual with the disease and the probability of incorrectly classifying an individual who does not have it. When the *BDT* is negative, the likelihood ratio, called the negative likelihood ratio $\left(LR^-\right)$, is the ratio between the probability of incorrectly classifying an individual who has the disease and the probability of correctly classifying an individual who does not have it. The *LRs* only depend on the sensitivity and the specificity of the *BDT* and do not depend on the disease prevalence, and therefore the *LRs* are superior parameters of the accuracy of a *BDT* (Zhou et al, 2011).

The comparison of the parameters of two *BDTs* has been the subject of numerous studies in Statistical literature. When the two *BDTs* and the *GS* are applied to all of the individuals in a random sample sized *n* (paired design), the comparison of the two sensitivities (specificities) is made by applying a comparison test of two paired binomial proportions. Subject to this same sample design, the comparison of the *LRs* of two *BDTs* is more complex. Leisenring and Pepe (1998) studied the estimation of the *LRs* of a *BDT* through a regression model. Pepe (2003) adapted this model to compare the *LRs* of two *BDTs*, for which in the regression model a variable dummy is considered to compare a *BDT* in relation to another. Moreover, Pepe proposed a confidence interval for the ratio of the two positive (negative) *LRs* estimating the variance of the ratios subject to the null hypothesis of equality of the two *LRs*. Section 3.1 summarizes the method of Pepe (2003). Biggerstaff (2000) proposed a graphical method to compare the *LRs* of two (or more) *BDTs*. Nevertheless, this method is not inferential and can only be applied to the estimators. Roldán-Nofuentes and Luna (2007) studied hypothesis tests to compare the *LRs* individually and simultaneously, and they also studied the same problem for the case of ordinal diagnostic tests. The hypothesis tests proposed by Roldán-Nofuentes and Luna (2007) are based on the logarithmic transformation of the ratio of the positive (negative) *LRs*, and therefore by inverting the test statistics of the individual tests, confidence intervals are obtained for the ratio of the two *LRs* (in Section 3.2 we summarize this method). Dolgun et al (2012) extended the method of Leisenring and Pepe (1998) to compare the *LRs* simultaneously.

Comparing the sensitivities (specificities) of two *BDTs*, we compare the intrinsic accuracy of both *BDTs*, and we determined which *BDT* is more accurate for an individual who has the disease (which *BDT* has the greatest sensitivity) or for an individual who does not have the disease (which *BDT* has the greatest specificity). Comparing the positive (negative) *LRs* of two *BDTs* it is possible to quantify with which *BDT* it is more likely to obtain a positive (negative) result for the *BDT* for an individual who has the disease than for an individual who does not.

In this manuscript we study the comparison of the *LRs* of two *BDTs* through confidence intervals (*CIs*), making the following contributions: a) four intervals to compare the *LRs*, and b) a method to calculate the sample size to compare the *LRs* through *CIs*. Section 2 presents the *LRs* and their properties. Section 3 presents the *CIs* studied by Pepe (2003), by Roldán-Nofuentes and Luna (2007), and four new *CIs* are proposed: a Wald type interval, an interval based on the Fieller method, a bootstrap interval based on the bias-corrected interval, and a Bayesian interval based on non-informative beta distributions and on the application of the Monte Carlo method. In Section 4, simulation experiments are carried out to study the coverage probabilities and the average lengths of the *CIs* presented in Section 3. Section 5 presents a method to calculate the sample size to compare the *LRs* through *CIs*. In Section 6, the results are applied to two real examples, and in Section 7 the results obtained are discussed.

## 2. Likelihood ratios

Let us consider a *BDT* that is assessed in relation to a *GS*. Let *T* be the variable that models the result of the *BDT*: $T=1$ when the *BDT* is positive and $T=0$ when it is negative. Let *D* be the variable that models the result of the *GS*: $D=1$ when the individual has the disease and $D=0$ when this is not the case. Let $\pi = P(D=1)$ be the disease prevalence in the population studied, and $\bar{\pi} = 1 - \pi$. The positive *LR* (Pepe, 2003; Zhou et al, 2011) is defined as

$$LR^+ = \frac{P(T=1|D=1)}{P(T=1|D=0)} = \frac{Se}{1-Sp}, \tag{1}$$

and the negative *LR* as

$$LR^{-} = \frac{P(T=0|D=1)}{P(T=0|D=0)} = \frac{1-Se}{Sp}. \qquad (2)$$

The *LRs* vary between 0 and infinity, and have the following properties:

a) If the *BDT* and the *GS* are independent then $LR^{+} = LR^{-} = 1$.

b) If the *BDT* correctly classifies all of the individuals then $LR^{+} = \infty$ and $LR^{-} = 0$.

c) If $LR^{+} > 1$ then a positive result in the *BDT* is more probable for an individual who has the disease than for an individual who does not.

d) If $LR^{-} < 1$ then a negative result in the *BDT* is more probable for an individual who does not have the disease than for an individual who does.

e) The *LRs* quantify the increase in knowledge of the presence of the disease through the application of the *BDT*. Before applying the test, the odds of an individual having the disease are pre-test odds $= \pi/(1-\pi)$, where $\pi$ is the disease prevalence. After applying the *BDT*, the odds are post-test odds $= \frac{P(D=1|T=i)}{P(D=0|T=i)}$, $i=0,1$. The *LRs* relate the pre-test odds and the post-test odds:

$$\text{post test odds } (T=1) = LR^{+} \times \text{pre test odds}$$
$$\text{post test odds } (T=0) = LR^{-} \times \text{pre test odds}.$$

Therefore, the likelihood ratios quantify the change in the odds of the disease obtained by knowledge of the application of the *BDT*.

We then study the comparison of the *LRs* of two *BDTs* subject to a paired design through *CIs*.

## 3. Confidence intervals

Let us consider two *BDTs* that are assessed in relation to the same *GS*. Let $T_h$ be the variable that models the result of the *h*th *BDT*, with $h = 1, 2$, defined in a similar way to the variable *T* given in Section 2. Let $Se_h$ and $Sp_h$ be the sensitivity and the specificity of the *h*th *BDT*, and $LR_h^+$ and $LR_h^-$ the positive and negative likelihood ratios respectively. Table 1 shows the frequencies and the theoretical probabilities obtained when comparing two *BDTs* in relation to a *GS* subject to a paired design. In the observed frequencies given in Table 1, the only value set by the researcher is the sample size *n*.

Table 1. Frequencies and probabilities subject to a paired design.

| | Frequencies | | | | |
|---|---|---|---|---|---|
| | $T_1 = 1$ | | $T_1 = 0$ | | |
| | $T_2 = 1$ | $T_2 = 0$ | $T_2 = 1$ | $T_2 = 0$ | Total |
| $D = 1$ | $s_{11}$ | $s_{10}$ | $s_{01}$ | $s_{00}$ | $s$ |
| $D = 0$ | $r_{11}$ | $r_{10}$ | $r_{01}$ | $r_{00}$ | $r$ |
| Total | $s_{11} + r_{11}$ | $s_{10} + r_{10}$ | $s_{01} + r_{01}$ | $s_{00} + r_{00}$ | $n$ |
| | Probabilities | | | | |
| | $T_1 = 1$ | | $T_1 = 0$ | | |
| | $T_2 = 1$ | $T_2 = 0$ | $T_2 = 1$ | $T_2 = 0$ | Total |
| $D = 1$ | $p_{11}$ | $p_{10}$ | $p_{01}$ | $p_{00}$ | $\pi$ |
| $D = 0$ | $q_{11}$ | $q_{10}$ | $q_{01}$ | $q_{00}$ | $\bar{\pi}$ |
| Total | $p_{11} + q_{11}$ | $p_{10} + q_{10}$ | $p_{01} + q_{01}$ | $p_{00} + q_{00}$ | 1 |

Applying the model of conditional dependence of Vacek (1985), the theoretical probabilities are expressed as

$$p_{ij} = \pi \left[ Se_1^i (1 - Se_1)^{1-i} Se_2^j (1 - Se_2)^{1-j} + \delta_{ij} \varepsilon_1 \right] \qquad (3)$$

and

$$q_{ij} = \bar{\pi}\left[ Sp_1^{1-i}(1-Sp_1)^i Sp_2^{1-j}(1-Sp_2)^j + \delta_{ij}\varepsilon_0 \right], \quad (4)$$

when $\delta_{ij} = 1$ if $i = j$ and $\delta_{ij} = -1$ if $i \neq j$, with $i, j = 0, 1$, and verifying that $\pi = \sum_{ij} p_{ij}$ and $\bar{\pi} = \sum_{ij} q_{ij}$. The parameters $\varepsilon_1$ and $\varepsilon_0$ are the dependence factors between the two *BDTs* when $D = 1$ and when $D = 0$ respectively, verifying that $0 \leq \varepsilon_1 \leq Min\{Se_1(1-Se_2), Se_2(1-Se_1)\}$ and $0 \leq \varepsilon_0 \leq Min\{Sp_1(1-Sp_2), Sp_2(1-Sp_1)\}$. If $\varepsilon_1 = \varepsilon_0 = 0$ then the two *BDTs* are conditionally independent from the disease, which is not normally a realistic one. In practice, the *BDTs* are conditionally dependent on the disease, so that $\varepsilon_1 > 0$ and/or $\varepsilon_0 > 0$. The frequencies of Table 1 are the product of a multinomial distribution whose vector of probabilities is $\boldsymbol{\psi} = (p_{11}, p_{10}, p_{01}, p_{00}, q_{11}, q_{10}, q_{01}, q_{00})^T$. The maximum likelihood estimators of these probabilities are $\hat{p}_{ij} = s_{ij}/n$ and $\hat{q}_{ij} = r_{ij}/n$, those of $\pi$ and $\bar{\pi}$ are $\hat{\pi} = s/n$ and $\hat{\bar{\pi}} = r/n$, and the variance-covariance matrix of $\hat{\boldsymbol{\psi}}$ is $\Sigma_{\hat{\boldsymbol{\psi}}} = \{\text{diag}(\boldsymbol{\psi}) - \boldsymbol{\psi}\boldsymbol{\psi}^T\}/n$.

In terms of the probabilities of the vector $\boldsymbol{\psi}$, the sensitivity and the specificity of each *BDT* are written as $Se_1 = (p_{10} + p_{11})/\pi$, $Sp_1 = (q_{00} + q_{01})/\bar{\pi}$, $Se_2 = (p_{01} + p_{11})/\pi$ and $Sp_2 = (q_{00} + q_{10})/\bar{\pi}$. The estimators of the sensitivities and the specificities are $\hat{Se}_1 = \frac{s_{11} + s_{10}}{s}$, $\hat{Se}_2 = \frac{s_{11} + s_{01}}{s}$, $\hat{Sp}_1 = \frac{r_{01} + r_{00}}{r}$ and $\hat{Sp}_2 = \frac{r_{10} + r_{00}}{r}$, and those of the dependence factors are $\hat{\varepsilon}_1 = \frac{\hat{p}_{11}}{\hat{\pi}} - \hat{Se}_1\hat{Se}_2 = \frac{s_{11}s_{00} - s_{10}s_{01}}{s}$ and $\hat{\varepsilon}_0 = \frac{\hat{q}_{00}}{\hat{\bar{\pi}}} - \hat{Sp}_1\hat{Sp}_2 = \frac{r_{11}r_{00} - r_{10}r_{01}}{r}$. Applying the delta method, it holds that the variances-covariances of $\hat{Se}_h$ and $\hat{Sp}_h$ are

$$Var\left(\hat{Se}_h\right) = \frac{Se_h(1-Se_h)}{n\pi}, \quad Var\left(\hat{Sp}_h\right) = \frac{Sp_h(1-Sp_h)}{n\bar{\pi}},$$
$$Cov\left(\hat{Se}_1, \hat{Se}_2\right) = \frac{\varepsilon_1}{n\pi}, \quad Cov\left(\hat{Sp}_1, \hat{Sp}_2\right) = \frac{\varepsilon_0}{n\bar{\pi}}. \quad (5)$$

The rest of the covariances are zero. Regarding the *LRs*, applying the delta method again, their variances-covariances (the proof can be seen in Appendix A) are

$$Var\left(\hat{LR}_h^+\right) \approx \frac{Se_h^2 Var\left(\hat{Sp}_h\right) + (1-Sp_h)^2 Var\left(\hat{Se}_h\right)}{(1-Sp_h)^4},$$

$$Var\left(\hat{LR}_h^-\right) \approx \frac{(1-Se_h)^2 Var\left(\hat{Sp}_h\right) + Sp_h^2 Var\left(\hat{Se}_h\right)}{Sp_h^4},$$

$$Cov\left(\hat{LR}_1^+, \hat{LR}_1^+\right) \approx \frac{Se_1 Se_2 Cov\left(\hat{Sp}_1, \hat{Sp}_2\right) + (1-Sp_1)(1-Sp_2) Cov\left(\hat{Se}_1, \hat{Se}_2\right)}{(1-Sp_1)^2 (1-Sp_2)^2}, \quad (6)$$

$$Cov\left(\hat{LR}_1^-, \hat{LR}_1^-\right) \approx \frac{(1-Se_1)(1-Se_2) Cov\left(\hat{Sp}_1, \hat{Sp}_2\right) + Sp_1 Sp_2 Cov\left(\hat{Se}_1, \hat{Se}_2\right)}{Sp_1^2 Sp_2^2}.$$

Substituting in the previous expressions the parameters with their estimators, we obtain the expressions of the estimators of the variances-covariances. Pepe (2003) studied the comparison of the *LRs* considering the ratio between them, i.e. $\omega^+ = LR_1^+/LR_2^+$ and $\omega^- = LR_1^-/LR_2^-$. Roldán-Nofuentes and Luna (2007) considered the Napierian logarithm of $\omega$. In this study, we are going to follow the same criteria as Pepe, and therefore we are going to compare the *LRs* through *CIs* for $\omega^+$ and $\omega^-$. From here onwards, we are going to consider that $LR_h$ is $LR_h^+$ or $LR_h^-$, and that $\omega$ is $\omega^+$ or $\omega^-$, depending on whether we compare the positive *LRs* or the negative *LRs*. If the *CI* for $\omega$ contains the value one, then we do not reject the equality of the *LRs* of both *BDTs*; in the opposite case, the *LR* of a *BDT* is significantly higher than that of the other *BDT*. Applying the delta method (see Appendix A), the variance of $\hat{\omega}$ is

$$Var(\hat{\omega}) \approx \omega^2 \left[ \frac{Var(\hat{LR}_1)}{LR_1^2} + \frac{Var(\hat{LR}_2)}{LR_2^2} - \frac{2Cov(\hat{LR}_1, \hat{LR}_2)}{LR_1 LR_2} \right]. \quad (7)$$

Then six *CIs* are presented for each ratio $\omega^+$ and $\omega^-$. The first interval was proposed by Pepe (2003), the second is deduced from the study by Roldán-Nofuentes and Luna (2007), and the rest of the intervals are contributions made by this manuscript.

*3.1. Regression model*

Leisenring and Pepe (1998) studied the estimation of the *LRs* of a *BDT* in presence of covariates through a regression model. For the positive *LR*, the regression model with $p$ covariates is $\ln(LR^+(X_1)) = \beta_0 + \sum_{i=1}^{p} \beta_i X_{1p}$, where $\beta_i$ are the parameters of the model and $X_1 = (X_{11}, ..., X_{1p})$ is the matrix of covariates. This model can be used to compare two *BDTs* (Pepe, 2003), i.e. $\ln[LR^+(X_T)] = \beta_0 + \beta_1 X_T$, where $X_T$ is a variable dummy to compare a *BDT* in relation to another. The regression model to compare the two negative *LRs* is $\ln[LR^+(X_T)] = \alpha_0 + \alpha_1 X_T$. In these models, the ratio $\omega^+$ is estimated as $e^{\hat{\beta}_1}$ and the ratio $\omega^-$ as $e^{\hat{\alpha}_1}$. The confidence interval for $\omega^+$ is

$$\hat{\omega}^+ \times \exp\left\{\pm z_{1-\alpha/2} \sqrt{\hat{Var}_0\left[\ln(\hat{\omega}^+)\right]}\right\}, \quad (8)$$

where $z_{1-\alpha/2}$ is the $100(1-\alpha/2)th$ percentile of the standard normal distribution and

$$\hat{Var}_0\left[\ln(\hat{\omega}^+)\right] \approx \frac{1-\hat{Se}_1}{s\hat{Se}_1} + \frac{\hat{Sp}_1}{r(1-\hat{Sp}_1)} + \frac{1-\hat{Se}_2}{s\hat{Se}_2} + \frac{\hat{Sp}_2}{r(1-\hat{Sp}_2)}$$

is the estimated variance of $\hat{\omega}^+$ subject to the null hypothesis $H_0: LR_1^+ = LR_2^+$. The confidence interval for $\omega^-$ s similar to the previous one, where

$$\hat{Var}_0\left[\ln(\hat{\omega}^-)\right] \approx \frac{\hat{Se}_1}{s(1-\hat{Se}_1)} + \frac{1-\hat{Sp}_1}{r\hat{Sp}_1} + \frac{\hat{Se}_1}{s(1-\hat{Se}_1)} + \frac{1-\hat{Sp}_1}{r\hat{Sp}_1}.$$

The book by Pepe (2003) discusses the confidence interval obtained from the regression model.

*3.2. Logarithmic interval*

Roldán-Nofuentes and Luna (2007) studied a hypothesis test to compare the positive (negative) *LRs* of two *BDTs* subject to a paired design. These hypothesis tests are based on the transformation of the Napierian logarithm of the ratio between the two positive (negative) *LRs*, i.e., $H_0: \ln(\omega)=0$ vs $H_1: \ln(\omega) \neq 0$, where $\omega$ is $\omega^+ = LR_1^+/LR_2^+$ or $\omega^- = LR_1^-/LR_2^-$, and the test statistic is

$$\frac{\ln(\hat{\omega})}{\sqrt{\hat{Var}\left[\ln(\hat{\omega})\right]}} \to N(0,1), \qquad (9)$$

where $\hat{Var}\left[\ln(\hat{\omega})\right]$ is an unrestricted estimator of the variance and is calculated applying the delta method (see Appendix A), i.e.

$$Var\left[\ln(\hat{\omega})\right] \approx \frac{Var(\hat{LR}_1)}{LR_1^2} + \frac{Var(\hat{LR}_2)}{LR_2^2} - \frac{2Cov(\hat{LR}_1, \hat{LR}_2)}{LR_1 LR_2}, \qquad (10)$$

and substituting in this expression each parameter with its estimator. Inverting the test statistic (9), it holds that the *CI* for $\ln(\omega)$ is $\ln(\hat{\omega}) \pm z_{1-\alpha/2}\sqrt{\hat{Var}\left[\ln(\hat{\omega})\right]}$. Finally, the logarithmic *CI* for $\omega$ is

$$\hat{\omega} \times \exp\left\{\pm z_{1-\alpha/2} \sqrt{\hat{V}ar\left[\ln(\hat{\omega})\right]}\right\}. \tag{11}$$

Roldán-Nofuentes and Luna studied the size (and the power) of the test $H_0 : \ln(\omega) = 0$ through simulation experiments. As the logarithmic interval (11) is obtained by inverting the test statistic (9), the coverage probability of this interval is equal to 1 minus the type I error obtained in the simulations carried out by Roldán-Nofuentes and Luna, and therefore the results are equivalent.

### 3.3. Wald CI

The Wald interval (Wald, 1943) is a classic interval for a parameter. Assuming the asymptotic normality of $\hat{\omega}$, i.e. $\hat{\omega} \xrightarrow[n\to\infty]{} N\left[\omega, \text{Var}(\omega)\right]$, the Wald CI for $\omega$ is

$$\hat{\omega}\left[1 \pm z_{1-\alpha/2} \sqrt{\frac{\hat{V}ar(\hat{LR}_1)}{\hat{LR}_1^2} + \frac{\hat{V}ar(\hat{LR}_2)}{\hat{LR}_2^2} - \frac{2\hat{C}ov(\hat{LR}_1, \hat{LR}_2)}{\hat{LR}_1\hat{LR}_2}}\right]. \tag{12}$$

### 3.4. Fieller CI

The Fieller method (1940) is a classic method used to calculate a *CI* for the ratio of two parameters, and requires us to assume that the estimators are distributed according to a normal bivariate distribution. Therefore, assuming the bivariant normality, i.e. $\left(\hat{LR}_1, \hat{LR}_2\right)^T \xrightarrow[n\to\infty]{} N\left[(LR_1, LR_2)^T, \Sigma_{\mathbf{LR}}\right]$, where

$$\Sigma_{\mathbf{LR}} = \begin{pmatrix} Var(LR_1) & Cov(LR_1, LR_2) \\ Cov(LR_1, LR_2) & Var(LR_2) \end{pmatrix},$$

and applying the Fieller method, it is verified that

$$\hat{LR}_1 - \omega\hat{LR}_2 \xrightarrow[n\to\infty]{} N\left(0, Var(LR_1) - 2\omega Cov(LR_1, LR_2) + \omega^2 Var(LR_2)\right).$$

The Fieller *CI* is obtained by searching for the set of values for $\omega$ that satisfy the inequality

$$\frac{\left(\hat{LR}_1 - \omega\hat{LR}_2\right)^2}{\hat{Var}\left(\hat{LR}_1\right) - 2\omega\hat{Cov}\left(\hat{LR}_1, \hat{LR}_2\right) + \omega^2\hat{Var}\left(\hat{LR}_2\right)} < z^2_{1-\alpha/2}.$$

Solving this inequation, the Fieller *CI* for $\omega$ is

$$\frac{\hat{LR}_1\hat{LR}_2 - \hat{\sigma}_{12}z^2_{1-\alpha/2} \pm \sqrt{\left(\hat{LR}_1\hat{LR}_2 - \hat{\sigma}_{12}z^2_{1-\alpha/2}\right)^2 - \left(\hat{LR}_1^2 - \hat{\sigma}_{11}z^2_{1-\alpha/2}\right)\left(\hat{LR}_2^2 - \hat{\sigma}_{22}z^2_{1-\alpha/2}\right)}}{\left(\hat{LR}_2^2 - \hat{\sigma}_{22}z^2_{1-\alpha/2}\right)}, \quad (13)$$

where $\hat{\sigma}_{ii} = \hat{Var}\left(\hat{LR}_i\right)$ and $\hat{\sigma}_{12} = \hat{Cov}\left(\hat{LR}_1, \hat{LR}_2\right)$. This interval is valid when $\left(\hat{LR}_1\hat{LR}_2 - \hat{\sigma}_{12}z^2_{1-\alpha/2}\right)^2 > \left(\hat{LR}_1^2 - \hat{\sigma}_{11}z^2_{1-\alpha/2}\right)\left(\hat{LR}_2^2 - \hat{\sigma}_{22}z^2_{1-\alpha/2}\right)$ and $\hat{LR}_2^2 - \hat{\sigma}_{22}z^2_{1-\alpha/2} \neq 0$.

*3.5. Bootstrap CI*

The Bootstrap method is one which is widely used for the estimation of parameters. The Bootstrap *CI* is calculated generating *B* random samples with replacement from the sample sized *n*, and then a *CI* is calculated. For the interval, we considered the bias-corrected Bootstrap *CI* (Efron and Tibshirani, 1993). For each one of the *B* samples with replacement, we calculate the estimators of the *LRs* and of $\omega$, i.e. $\hat{LR}_{1Bi}$, $\hat{LR}_{2Bi}$ and $\hat{\omega}_{Bi}$, with $i = 1,...,B$. The parameter $\omega$ is estimated as the average of the *B* Bootstrap estimations, i.e. $\hat{\bar{\omega}}_B = \frac{1}{B}\sum_{i=1}^{B}\hat{\omega}_{Bi}$. Let $A = \#\left(\hat{\omega}_{Bi} < \hat{\omega}\right)$ be the number of samples in which the Bootstrap estimator $\hat{\omega}_{Bi}$ is lower than the maximum likelihood estimator $\hat{\omega}$. Let $\hat{z}_0 = \Phi^{-1}(A/B)$, where $\Phi^{-1}(\cdot)$ is the inverse function of the standard normal cumulative distribution function. Let $q_1 = \Phi(2\hat{z}_0 - z_{1-\alpha/2})$ and $q_2 = \Phi(2\hat{z}_0 + z_{1-\alpha/2})$, then the bias-corrected Bootstrap *CI* is

$$\left(\hat{\omega}_B^{(q_1)},\ \hat{\omega}_B^{(q_2)}\right) \tag{14}$$

where $\hat{\omega}_B^{(q)}$ is the *q*th quantile of the distribution of the *B* Bootstrap estimations of $\omega$. The bias-corrected bootstrap *CI* is consistent, as it verifies (Shao and Tu, 1995) that $P\left[\sqrt{n}(\hat{\omega}_n - \omega) \leq x\right] - P_B\left[\sqrt{n}(\hat{\omega}_{B,n} - \hat{\omega}_n) \leq x\right]$ converges in probability to zero when the sample size is very large $(n \to \infty)$ for every value *x*, where $P_B$ is the bootstrap distribution and $\hat{\omega}_{B,n}$ is the is the upper (lower) limit of the bootstrap *CI*.

### 3.6. Bayesian CI

The previous *CIs* are all frequentists, the problem can also be addressed from a Bayesian perspective. Conditioning on $D=1$, i.e. on the individuals who have the disease, it is verified that $s_{11} + s_{10} \to B(s, Se_1)$ and that $s_{11} + s_{01} \to B(s, Se_2)$. Conditioning on $D=0$ it is verified that $r_{01} + r_{00} \to B(r, Sp_1)$ and that $r_{10} + r_{00} \to B(r, Sp_2)$. Considering the distribution of the *BDT* 1, the estimators of its sensitivity and specificity are $\hat{Se}_1 = \dfrac{s_{11} + s_{10}}{s}$ and $\hat{Sp}_1 = \dfrac{r_{01} + r_{00}}{r}$, which are estimators of binomial proportions. In a similar way, the estimators $\hat{Se}_2 = \dfrac{s_{11} + s_{01}}{s}$ and $\hat{Sp}_2 = \dfrac{r_{10} + r_{00}}{r}$ are also estimators of binomial proportions. Therefore, for these estimators, conjugate beta prior distributions are proposed, i.e.

$$\hat{Se}_h \to Beta(\alpha_{Se_h}, \beta_{Se_h}) \text{ and } \hat{Sp}_h \to Beta(\alpha_{Sp_h}, \beta_{Sp_h}), \tag{15}$$

with $h = 1, 2$. Let $\mathbf{n} = (s_{11}, s_{10}, s_{01}, s_{00}, r_{11}, r_{10}, r_{01}, r_{00})$ be the vector of observed frequencies, then the posteriori distributions for the estimators of the sensitivity and the specificity of the *BDT* 1 are

$$\hat{S}e_1 | \mathbf{n} \rightarrow Beta\left(s_{11} + s_{10} + \alpha_{Se_1}, s_{01} + s_{00} + \beta_{Se_1}\right) \tag{16}$$

and

$$\hat{S}p_1 | \mathbf{n} \rightarrow Beta\left(r_{01} + r_{00} + \alpha_{Sp_1}, r_{11} + r_{10} + \beta_{Sp_1}\right). \tag{17}$$

In a similar way, the posteriori distributions for the estimators of the sensitivity and the specificity of the *BDT* 2 are

$$\hat{S}e_2 | \mathbf{n} \rightarrow Beta\left(s_{11} + s_{01} + \alpha_{Se_2}, s_{10} + s_{00} + \beta_{Se_2}\right) \tag{18}$$

and

$$\hat{S}p_2 | \mathbf{n} \rightarrow Beta\left(r_{10} + r_{00} + \alpha_{Sp_2}, r_{11} + r_{01} + \beta_{Sp_2}\right). \tag{19}$$

Once all the distributions have been defined, the posteriori distribution for the *LRs* of each *BDT*, and for $\omega^+$ and $\omega^-$, can be approximated by applying the Monte Carlo method (Boos and Stefanski, 2013). This method consists of generating $M$ random values of the posteriori distributions given in equations (16) to (19). In each interaction the generated values of sensitivities $\left(\hat{S}e_{hi}\right)$ and specificities $\left(\hat{S}p_{hi}\right)$ are plugged in the equations $\hat{L}R_{hi}^+ = \dfrac{\hat{S}e_{hi}}{1 - \hat{S}p_{hi}}$ and $\hat{L}R_{hi}^- = \dfrac{1 - \hat{S}e_{hi}}{\hat{S}p_{hi}}$, and from these each ratio $\hat{\omega}_i$ is calculated. As an estimator of each ratio the average of the $M$ Bayesian estimations is calculated, i.e. $\hat{\bar{\omega}}_{Ba} = \dfrac{1}{M} \sum_{i=1}^{M} \hat{\omega}_i$. Finally, from the $M$ values $\hat{\omega}_i$ a *CI* based on the quantiles is calculated, i.e. the $100 \times (1 - \alpha)\%$ *CI* for $\omega$ is

$$\left( \hat{\omega}_{Ba}^{(\alpha/2)}, \hat{\omega}_{Ba}^{(1-\alpha/2)} \right), \tag{20}$$

where $\hat{\omega}_{Ba}^{(q)}$ is the $q$th quantile of the distribution of the $M$ Bayesian estimations $\hat{\omega}_i$.

All of the *CIs* presented are for $\omega = LR_1/LR_2$. If we want to calculate the *CI* for $LR_2/LR_1$ $(=\omega' = 1/\omega)$, the regression, logarithmic, Fieller, Bootstrap and Bayesian intervals are obtained by calculating the inverse of each boundary of the corresponding interval for $\omega$. Nevertheless, the Wald *CI* for $\omega'$ is obtained from the Wald *CI* for $\omega$ dividing each boundary by $\hat{\omega}^2$, i.e. if $(L_\omega, U_\omega)$ is the Wald *CI* for $\omega$ then the Wald *CI* for $\omega' = 1/\omega$ is $(L_\omega/\hat{\omega}^2, U_\omega/\hat{\omega}^2)$.

## 4. Simulation experiments

Monte Carlo simulation experiments were carried out to study the coverage probability (*CP*) and the average length (*AL*) of each one of the *CIs* presented in the Section 3. For this purpose, $N = 10,000$ random samples of multinomial distributions with sizes $n = \{50, 100, 200, 300, 400, 500, 1000\}$ were generated, and their probabilities were calculated from equations (3) and (4). As the sensitivity and the specificity of each *BDT*, the values $Se_h, Sp_h = \{0.70, 0.75, ..., 0.90, 0.95\}$ were taken, which are realistic values in clinical practice, and the *LRs* were calculated with the equations $LR_h^+ = Se_h/(1-Sp_h)$ and $LR_h^- = (1-Se_h)/Sp_h$ with $h = 1, 2$. For the disease prevalence, $\pi = \{10\%, 25\%, 50\%\}$ was considered, and for the dependence factors $\varepsilon_1$ and $\varepsilon_0$ intermediate values (50% of the maximum value of each $\varepsilon_i$) and high values (80% of the maximum value of each $\varepsilon_i$) were taken, i.e.

$$\varepsilon_1 = k \times Min\{Se_1(1-Se_2), Se_2(1-Se_1)\} \text{ and } \varepsilon_0 = k \times Min\{Sp_1(1-Sp_2), Sp_2(1-Sp_1)\},$$

where $k = \{0.50, 0.80\}$. Once the value of the parameters in each scenario was set, the probabilities of each multinomial distribution were calculated by substituting the value of the parameters in equations (3) and (4).

For the Bootstrap interval, for each one of the $N$ random samples generated, $B = 2,000$ replacement samples were generated in turn, and from the $B$ replacement samples the bias-corrected bootstrap $CI$ was calculated through the method described in Section 3.5.

Regarding the Bayesian $CI$, for the estimators of the two sensitivities and of the two specificities, the $Beta(1,1)$ distribution was considered as prior distribution. The choice of this distribution is justified by the fact that it is a non-informative distribution, which is flat for every possible value of the sensitivities and the specificities, and it has a minimum impact on the posteriori distributions. Moreover, for each one of the $N$ generated random samples, $M = 10,000$ random samples were generated in turn, and from the $M$ samples the Bayesian $CI$ was been calculated by applying the method described in Section 3.6.

The simulation experiments were designed so that in every random sample generated, it is possible to estimate all the parameters and their variances-covariances. Therefore, if a parameter could not be estimated in a sample (for example, $\hat{Se}_h = 0$) then that sample was discarded and another one was generated in its place. This problem mainly occurred in the samples with a size of 50. In each one of the scenarios considered (values set for $Se_h$, $Sp_h$, $\pi$, $\varepsilon_1$ and $\varepsilon_0$) the coverage probability (*CP*) and the average length (*AL*) were calculated for each one of the six *CIs* for $\omega^+$ and $\omega^-$. The *CP* of each *CI* was calculated as the quotient between the number of intervals that contained the parameter ($\omega^+$ or $\omega^-$, depending on the case) and the number of samples generated *N*, and the *AL* was calculated adding the length of the *N* intervals and dividing this number by *N*. As the confidence level we took 95%.

The comparison of the asymptotic behaviour of the *CIs* was made following the criterion based on whether the *CI* "fails" or "does not fail" for a confidence of 95%. This criterion, which has been used by other authors (Price and Bonett, 2004; Martín-Andrés and Álvarez-Hernández, 2014a, 2014b; Montero-Alonso and Roldán-Nofuentes, 2018), establishes that a *CI* fails (or does not fail) if its coverage probability is $\leq 93\%$ $(>93\%)$. The selection of the *CI* with the best asymptotic behaviour was made through the following steps: 1) Choose the *CIs* with the fewest failures, and 2) Choose the *CIs* which are the most accurate, i.e. those with least *AL*, and among these those which have a *CP* closest to 95%. This method is justified in Appendix *B*.

*4.1. Positive LRs*

Tables 2 and 3 show some of the results obtained for the intervals of $\omega^+$, considering two different scenarios of sensitivities and specificities. In these tables, failures are indicated in bold type. From the results of the experiments, the following conclusions are reached:

   *a) Regression method*. The *CI* obtained applying the regression method does not fail, and it has a *CP* of 100% or very close to this value. In general terms, its *AL* is larger than that of the rest of the intervals.

*b) Logarithmic CI.* The logarithmic *CI* does not fail. In very general terms, when the sample size is small $(n = 50)$ or moderate $(n = 100)$ its *CP* is 100% or very near to this value. When the sample size is large $(n = 200 - 400)$ or very large $(n \geq 500)$ its *CP* fluctuates around 95%. The *AL* of this interval is lower than that of the interval calculated through regression.

*c) Wald CI.* When $\omega^+ \neq 1$, this interval may fail if $n \leq 100$ and the prevalence is moderate $(\pi = 25\%)$ or large $(\pi = 50\%)$, whereas if $n \geq 200$ the interval does not fail. When $\omega^+ = 1$ the interval does not fail. In situations in which the Wald *CI* does not fail, its *CP* and *AL* are very similar to those of the logarithmic *CI*.

*c) Fieller CI.* The Fieller *CI* does not fail. In general terms, its *CP* is 100% or very close to this value when $n \leq 100$. When $n \geq 200$ its *CP* behaves in a very similar way to the *CP* of the logarithmic and Wald intervals (and the *ALs* are very similar). Therefore, when $n \geq 200$, the behaviour of the Fieller *CI* is very similar to the logarithmic and Wald intervals.

*d) Bootstrap CI.* In very general terms, when $n \leq 100$ this interval may fail if $\omega^+ \neq 1$ or if it has a *CP* which is equal to or very near to 100% if $\omega^+ = 1$. When $n \geq 200$, the Bootstrap *CI* does not fail, its *CP* fluctuates around 95% and its *AL* is very similar to that of the logarithmic, Wald and Fieller intervals. Therefore, when $n \geq 200$ the Bootstrap interval has an asymptotic behaviour which is very similar to that of logarithmic, Wald and Fieller intervals.

Table 2. Coverage probabilities and average lengths of the *CIs* for the ratio of the two positive *LRs* (I).

$LR_1^+ = 9.5 \quad LR_2^+ = 4.5 \quad LR_1^- = 0.056 \quad LR_2^- = 0.125 \quad \omega^+ = 2.111 \quad \omega^- = 0.444$

$Se_1 = 0.95 \quad Sp_1 = 0.90 \quad Se_2 = 0.90 \quad Sp_2 = 0.80$

$\pi = 10\% \quad \varepsilon_1 = 0.0225 \quad \varepsilon_0 = 0.0400$

| | Regression | | Logarithmic | | Wald | | Fieller | | Bootstrap | | Bayesian | |
|---|---|---|---|---|---|---|---|---|---|---|---|---|
| n | CP | AL | CP | AL | CP | AL | CP | AL | CP | AL | CP | AL |
| 50 | 99.95 | 7.06 | 99.40 | 5.72 | 97.20 | 4.53 | 100 | 8.93 | 98.30 | 3.09 | 99.90 | 5.90 |
| 100 | 99.25 | 5.73 | 97.90 | 4.75 | 97.40 | 4.16 | 99.80 | 5.64 | 98.50 | 3.69 | 99.10 | 5.42 |
| 200 | 99.40 | 3.04 | 96.85 | 2.49 | 96.60 | 2.38 | 97.90 | 2.61 | 96.90 | 2.52 | 99.30 | 3.04 |
| 300 | 98.90 | 2.26 | 96.15 | 1.86 | 96.10 | 1.81 | 96.85 | 1.90 | 95.60 | 1.89 | 99.00 | 2.27 |
| 400 | 99.10 | 1.86 | 95.90 | 1.53 | 95.85 | 1.50 | 96.10 | 1.55 | 95.80 | 1.55 | 99.15 | 1.86 |
| 500 | 98.50 | 1.61 | 95.55 | 1.33 | 95.45 | 1.31 | 95.90 | 1.35 | 95.05 | 1.34 | 98.35 | 1.62 |
| 1000 | 98.20 | 1.07 | 95.45 | 0.89 | 95.30 | 0.88 | 95.65 | 0.89 | 95.35 | 0.90 | 98.20 | 1.08 |

$\pi = 10\% \quad \varepsilon_1 = 0.0360 \quad \varepsilon_0 = 0.0640$

| | Regression | | Logarithmic | | Wald | | Fieller | | Bootstrap | | Bayesian | |
|---|---|---|---|---|---|---|---|---|---|---|---|---|
| n | CP | AL | CP | AL | CP | AL | CP | AL | CP | AL | CP | AL |
| 50 | 99.95 | 6.54 | 99.10 | 4.78 | 95.50 | 3.93 | 99.95 | 7.74 | 91.80 | 2.51 | 99.95 | 5.48 |
| 100 | 99.90 | 5.15 | 98.60 | 3.76 | 96.55 | 3.39 | 99.45 | 4.57 | 95.60 | 2.72 | 99.90 | 4.91 |
| 200 | 99.60 | 2.93 | 96.90 | 2.09 | 96.00 | 2.01 | 98.15 | 2.19 | 96.35 | 1.95 | 99.55 | 2.93 |
| 300 | 99.65 | 2.21 | 96.30 | 1.57 | 95.90 | 1.53 | 97.25 | 1.61 | 96.00 | 1.53 | 99.60 | 2.22 |
| 400 | 99.80 | 1.82 | 95.90 | 1.30 | 95.95 | 1.28 | 97.10 | 1.32 | 96.30 | 1.28 | 99.85 | 1.83 |
| 500 | 99.75 | 1.59 | 95.80 | 1.13 | 95.75 | 1.12 | 96.35 | 1.15 | 95.65 | 1.13 | 99.80 | 1.60 |
| 1000 | 99.55 | 1.07 | 95.45 | 0.76 | 95.35 | 0.76 | 95.70 | 0.77 | 95.50 | 0.76 | 99.60 | 1.08 |

$\pi = 25\% \quad \varepsilon_1 = 0.0225 \quad \varepsilon_0 = 0.0400$

| | Regression | | Logarithmic | | Wald | | Fieller | | Bootstrap | | Bayesian | |
|---|---|---|---|---|---|---|---|---|---|---|---|---|
| n | CP | AL | CP | AL | CP | AL | CP | AL | CP | AL | CP | AL |
| 50 | 99.85 | 6.04 | 97.80 | 4.89 | **91.30** | 3.95 | 99.90 | 6.38 | 93.60 | 2.72 | 99.65 | 5.49 |
| 100 | 99.50 | 5.19 | 97.90 | 4.28 | 95.05 | 3.74 | 99.40 | 4.52 | 97.45 | 3.28 | 99.35 | 4.90 |
| 200 | 98.45 | 2.96 | 95.60 | 2.44 | 94.75 | 2.32 | 97.30 | 2.50 | 95.90 | 2.62 | 98.40 | 2.91 |
| 300 | 98.45 | 2.28 | 95.45 | 1.88 | 95.25 | 1.83 | 97.05 | 1.91 | 94.95 | 2.03 | 98.30 | 2.25 |
| 400 | 99.00 | 1.91 | 96.10 | 1.59 | 95.95 | 1.55 | 96.65 | 1.60 | 95.60 | 1.68 | 98.85 | 1.90 |
| 500 | 98.55 | 1.65 | 95.60 | 1.37 | 95.25 | 1.35 | 96.15 | 1.38 | 95.55 | 1.43 | 98.55 | 1.65 |
| 1000 | 98.30 | 1.14 | 95.15 | 0.95 | 94.90 | 0.94 | 95.30 | 0.95 | 94.65 | 0.97 | 98.35 | 1.14 |

$\pi = 25\% \quad \varepsilon_1 = 0.0360 \quad \varepsilon_0 = 0.0640$

| | Regression | | Logarithmic | | Wald | | Fieller | | Bootstrap | | Bayesian | |
|---|---|---|---|---|---|---|---|---|---|---|---|---|
| n | CP | AL | CP | AL | CP | AL | CP | AL | CP | AL | CP | AL |
| 50 | 100 | 5.77 | 96.80 | 4.21 | **91.50** | 3.50 | 99.65 | 5.56 | **83.55** | 2.20 | 100 | 5.25 |
| 100 | 99.85 | 4.45 | 95.40 | 3.19 | **91.85** | 2.88 | 97.15 | 3.45 | **89.15** | 2.31 | 99.80 | 4.25 |
| 200 | 99.60 | 2.85 | 96.15 | 2.02 | 94.00 | 1.95 | 96.40 | 2.08 | 94.85 | 1.93 | 99.60 | 2.80 |
| 300 | 99.40 | 2.23 | 94.15 | 1.59 | 94.10 | 1.55 | 95.15 | 1.62 | 94.10 | 1.60 | 99.40 | 2.21 |
| 400 | 99.55 | 1.87 | 94.95 | 1.32 | 94.85 | 1.30 | 95.15 | 1.34 | 94.65 | 1.35 | 99.50 | 1.85 |
| 500 | 99.15 | 1.66 | 94.85 | 1.18 | 94.75 | 1.16 | 95.70 | 1.19 | 95.05 | 1.21 | 99.15 | 1.65 |
| 1000 | 99.50 | 1.14 | 95.00 | 0.81 | 95.15 | 0.81 | 95.70 | 0.82 | 94.90 | 0.83 | 99.30 | 1.14 |

$\pi = 50\% \quad \varepsilon_1 = 0.0225 \quad \varepsilon_0 = 0.0400$

| | Regression | | Logarithmic | | Wald | | Fieller | | Bootstrap | | Bayesian | |
|---|---|---|---|---|---|---|---|---|---|---|---|---|
| n | CP | AL | CP | AL | CP | AL | CP | AL | CP | AL | CP | AL |
| 50 | 99.75 | 5.98 | 96.75 | 4.88 | **89.35** | 3.80 | 99.75 | 6.11 | **86.45** | 2.31 | 99.55 | 5.39 |
| 100 | 99.60 | 5.91 | 96.35 | 4.87 | **92.20** | 3.97 | 98.90 | 5.22 | 94.45 | 2.81 | 99.40 | 5.38 |
| 200 | 98.85 | 3.78 | 95.90 | 3.13 | 94.15 | 2.89 | 97.70 | 3.21 | 96.85 | 3.10 | 98.70 | 3.65 |
| 300 | 98.50 | 2.87 | 95.00 | 2.38 | 94.70 | 2.26 | 96.40 | 2.41 | 95.40 | 2.61 | 98.30 | 2.82 |
| 400 | 98.50 | 2.40 | 95.35 | 1.99 | 95.05 | 1.92 | 96.80 | 2.02 | 94.65 | 2.20 | 98.25 | 2.37 |
| 500 | 98.35 | 2.08 | 95.80 | 1.72 | 95.45 | 1.68 | 95.25 | 1.74 | 95.25 | 1.88 | 98.20 | 2.06 |
| 1000 | 97.50 | 1.41 | 94.55 | 1.17 | 94.80 | 1.15 | 95.50 | 1.17 | 93.80 | 1.22 | 97.60 | 1.40 |

$\pi = 50\% \quad \varepsilon_1 = 0.0360 \quad \varepsilon_0 = 0.0640$

| | Regression | | Logarithmic | | Wald | | Fieller | | Bootstrap | | Bayesian | |
|---|---|---|---|---|---|---|---|---|---|---|---|---|
| n | CP | AL | CP | AL | CP | AL | CP | AL | CP | AL | CP | AL |
| 50 | 99.90 | 5.47 | 94.15 | 4.03 | **88.70** | 3.28 | 99.20 | 5.26 | **67.35** | 1.89 | 99.80 | 4.97 |
| 100 | 99.85 | 5.20 | 93.80 | 3.80 | **91.40** | 3.22 | 96.65 | 4.24 | **78.55** | 2.13 | 99.75 | 4.79 |
| 200 | 99.70 | 3.45 | 93.75 | 2.47 | 93.65 | 2.32 | 93.70 | 2.56 | **89.75** | 2.15 | 99.45 | 3.34 |
| 300 | 99.55 | 2.72 | 94.65 | 1.93 | 94.45 | 1.86 | 94.65 | 1.98 | 94.10 | 1.90 | 99.55 | 2.67 |
| 400 | 99.65 | 2.33 | 95.15 | 1.66 | 94.90 | 1.62 | 95.45 | 1.69 | 95.35 | 1.69 | 99.65 | 2.31 |
| 500 | 99.45 | 2.06 | 95.55 | 1.46 | 95.15 | 1.43 | 95.25 | 1.48 | 96.00 | 1.51 | 99.20 | 2.04 |
| 1000 | 99.20 | 1.40 | 94.75 | 1.00 | 94.80 | 0.99 | 94.85 | 1.00 | 94.80 | 1.03 | 99.25 | 1.40 |

Table 3. Coverage probabilities and average lengths of the *CIs* for the ratio of the two positive *LRs* (II).

$LR_1^+ = 6 \quad LR_2^+ = 6 \quad LR_1^- = 0.118 \quad LR_2^- = 0.118 \quad \omega^+ = 1 \quad \omega^- = 1$

$Se_1 = 0.90 \quad Sp_1 = 0.85 \quad Se_2 = 0.90 \quad Sp_2 = 0.85$

$\pi = 10\% \quad \varepsilon_1 = 0.0450 \quad \varepsilon_0 = 0.0638$

| | Regression | | Logarithmic | | Wald | | Fieller | | Bootstrap | | Bayesian | |
|---|---|---|---|---|---|---|---|---|---|---|---|---|
| n | CP | AL | CP | AL | CP | AL | CP | AL | CP | AL | CP | AL |
| 50 | 99.95 | 3.61 | 99.50 | 2.51 | 99.85 | 2.18 | 100 | 4.67 | 100 | 1.96 | 99.95 | 3.16 |
| 100 | 99.80 | 2.38 | 97.75 | 1.65 | 97.90 | 1.52 | 98.85 | 2.37 | 98.60 | 1.51 | 99.75 | 2.33 |
| 200 | 99.65 | 1.33 | 96.40 | 0.92 | 96.90 | 0.89 | 97.65 | 1.02 | 97.00 | 0.91 | 99.60 | 1.35 |
| 300 | 99.65 | 1.00 | 96.25 | 0.70 | 96.45 | 0.68 | 97.90 | 0.74 | 96.75 | 0.69 | 99.70 | 1.01 |
| 400 | 99.65 | 0.84 | 95.60 | 0.58 | 96.00 | 0.58 | 96.95 | 0.61 | 96.10 | 0.58 | 99.65 | 0.84 |
| 500 | 99.50 | 0.72 | 95.30 | 0.51 | 95.70 | 0.50 | 96.35 | 0.52 | 95.70 | 0.51 | 99.60 | 0.73 |
| 1000 | 99.25 | 0.48 | 94.65 | 0.34 | 94.30 | 0.34 | 95.15 | 0.35 | 94.80 | 0.34 | 99.25 | 0.49 |

$\pi = 10\% \quad \varepsilon_1 = 0.0720 \quad \varepsilon_0 = 0.1020$

| | Regression | | Logarithmic | | Wald | | Fieller | | Bootstrap | | Bayesian | |
|---|---|---|---|---|---|---|---|---|---|---|---|---|
| n | CP | AL | CP | AL | CP | AL | CP | AL | CP | AL | CP | AL |
| 50 | 100 | 3.18 | 100 | 1.79 | 99.90 | 1.62 | 100 | 3.65 | 100 | 1.43 | 100 | 2.77 |
| 100 | 100 | 2.19 | 99.85 | 1.11 | 99.75 | 1.06 | 100 | 1.58 | 99.95 | 0.99 | 100 | 2.15 |
| 200 | 100 | 1.28 | 98.15 | 0.60 | 98.20 | 0.59 | 98.75 | 0.67 | 98.55 | 0.57 | 100 | 1.29 |
| 300 | 100 | 0.98 | 97.05 | 0.45 | 97.15 | 0.45 | 97.45 | 0.48 | 97.95 | 0.43 | 100 | 0.98 |
| 400 | 100 | 0.82 | 96.85 | 0.37 | 96.90 | 0.37 | 97.05 | 0.39 | 97.15 | 0.37 | 100 | 0.82 |
| 500 | 100 | 0.71 | 96.30 | 0.33 | 96.40 | 0.32 | 96.80 | 0.34 | 96.65 | 0.32 | 100 | 0.72 |
| 1000 | 100 | 0.49 | 95.80 | 0.22 | 95.80 | 0.22 | 96.15 | 0.22 | 96.32 | 0.22 | 100 | 0.49 |

$\pi = 25\% \quad \varepsilon_1 = 0.0450 \quad \varepsilon_0 = 0.0638$

| | Regression | | Logarithmic | | Wald | | Fieller | | Bootstrap | | Bayesian | |
|---|---|---|---|---|---|---|---|---|---|---|---|---|
| n | CP | AL | CP | AL | CP | AL | CP | AL | CP | AL | CP | AL |
| 50 | 99.90 | 3.24 | 99.35 | 2.25 | 99.55 | 1.97 | 100 | 3.58 | 99.95 | 1.81 | 99.85 | 3.06 |
| 100 | 99.65 | 2.05 | 96.95 | 1.39 | 96.95 | 1.30 | 100 | 1.78 | 99.15 | 1.38 | 99.75 | 2.00 |
| 200 | 99.30 | 1.24 | 95.00 | 0.86 | 94.85 | 0.84 | 98.45 | 0.94 | 95.00 | 0.90 | 99.15 | 1.23 |
| 300 | 99.70 | 0.97 | 94.45 | 0.68 | 94.10 | 0.66 | 97.35 | 0.71 | 94.20 | 0.70 | 99.65 | 0.96 |
| 400 | 99.45 | 0.82 | 95.55 | 0.57 | 94.85 | 0.57 | 97.10 | 0.60 | 95.05 | 0.59 | 99.35 | 0.82 |
| 500 | 99.45 | 0.73 | 94.70 | 0.51 | 94.15 | 0.50 | 96.15 | 0.53 | 94.25 | 0.52 | 99.40 | 0.72 |
| 1000 | 99.60 | 0.51 | 95.45 | 0.36 | 95.25 | 0.36 | 95.85 | 0.36 | 95.15 | 0.36 | 99.50 | 0.51 |

$\pi = 25\% \quad \varepsilon_1 = 0.0720 \quad \varepsilon_0 = 0.1020$

| | Regression | | Logarithmic | | Wald | | Fieller | | Bootstrap | | Bayesian | |
|---|---|---|---|---|---|---|---|---|---|---|---|---|
| n | CP | AL | CP | AL | CP | AL | CP | AL | CP | AL | CP | AL |
| 50 | 100 | 2.80 | 100 | 1.49 | 99.85 | 1.38 | 100 | 2.51 | 100 | 1.27 | 100 | 2.66 |
| 100 | 100 | 1.93 | 99.30 | 0.89 | 99.25 | 0.86 | 100 | 1.15 | 100 | 0.82 | 100 | 1.89 |
| 200 | 100 | 1.21 | 96.95 | 0.53 | 96.50 | 0.53 | 98.70 | 0.59 | 98.30 | 0.53 | 100 | 1.20 |
| 300 | 100 | 0.96 | 95.85 | 0.42 | 95.65 | 0.42 | 96.75 | 0.45 | 97.65 | 0.42 | 100 | 0.95 |
| 400 | 100 | 0.82 | 95.35 | 0.36 | 94.95 | 0.36 | 96.30 | 0.38 | 96.35 | 0.37 | 100 | 0.82 |
| 500 | 100 | 0.73 | 95.25 | 0.32 | 95.25 | 0.32 | 95.90 | 0.33 | 95.80 | 0.33 | 100 | 0.73 |
| 1000 | 100 | 0.50 | 95.25 | 0.22 | 95.25 | 0.22 | 95.70 | 0.23 | 95.40 | 0.23 | 100 | 0.50 |

$\pi = 50\% \quad \varepsilon_1 = 0.0450 \quad \varepsilon_0 = 0.0638$

| | Regression | | Logarithmic | | Wald | | Fieller | | Bootstrap | | Bayesian | |
|---|---|---|---|---|---|---|---|---|---|---|---|---|
| n | CP | AL | CP | AL | CP | AL | CP | AL | CP | AL | CP | AL |
| 50 | 99.95 | 3.27 | 99.95 | 2.27 | 99.60 | 1.97 | 100 | 3.54 | 100 | 1.67 | 99.95 | 3.06 |
| 100 | 100 | 2.51 | 98.90 | 1.69 | 97.65 | 1.52 | 100 | 2.39 | 99.85 | 1.50 | 99.85 | 2.39 |
| 200 | 99.55 | 1.54 | 95.60 | 1.06 | 94.30 | 1.01 | 98.80 | 1.22 | 96.45 | 1.12 | 99.35 | 1.51 |
| 300 | 99.35 | 1.20 | 96.00 | 0.83 | 95.10 | 0.81 | 97.70 | 0.90 | 95.65 | 0.86 | 99.25 | 1.19 |
| 400 | 99.55 | 1.02 | 95.40 | 0.71 | 95.40 | 0.69 | 96.10 | 0.75 | 95.55 | 0.74 | 99.50 | 1.01 |
| 500 | 99.55 | 0.89 | 95.20 | 0.62 | 94.75 | 0.61 | 96.20 | 0.65 | 94.15 | 0.64 | 99.50 | 0.89 |
| 1000 | 99.55 | 0.61 | 94.40 | 0.43 | 94.75 | 0.43 | 95.75 | 0.44 | 94.25 | 0.44 | 99.50 | 0.61 |

$\pi = 50\% \quad \varepsilon_1 = 0.0720 \quad \varepsilon_0 = 0.1020$

| | Regression | | Logarithmic | | Wald | | Fieller | | Bootstrap | | Bayesian | |
|---|---|---|---|---|---|---|---|---|---|---|---|---|
| n | CP | AL | CP | AL | CP | AL | CP | AL | CP | AL | CP | AL |
| 50 | 100 | 2.81 | 100 | 1.50 | 99.95 | 1.38 | 100 | 2.58 | 100 | 1.24 | 100 | 2.66 |
| 100 | 100 | 2.25 | 99.90 | 1.05 | 99.70 | 1.00 | 100 | 1.51 | 100 | 0.94 | 100 | 2.16 |
| 200 | 100 | 1.49 | 99.20 | 0.66 | 98.45 | 0.65 | 99.95 | 0.77 | 99.95 | 0.64 | 100 | 1.47 |
| 300 | 100 | 1.17 | 97.70 | 0.51 | 97.05 | 0.50 | 99.50 | 0.56 | 99.45 | 0.51 | 100 | 1.16 |
| 400 | 100 | 1.00 | 96.50 | 0.43 | 96.40 | 0.43 | 98.55 | 0.46 | 97.95 | 0.44 | 100 | 0.99 |
| 500 | 100 | 0.89 | 95.75 | 0.39 | 95.35 | 0.38 | 97.55 | 0.40 | 96.80 | 0.39 | 100 | 0.88 |
| 1000 | 99.95 | 0.61 | 95.55 | 0.27 | 95.25 | 0.27 | 96.65 | 0.28 | 95.60 | 0.27 | 99.95 | 0.61 |

*e) Bayesian CI*. The Bayesian *CI* does not fail and has a *CP* and an *AL* which are very similar to those of the interval obtained by regression. The *CP* and the *AL* of the Bayesian interval are almost always higher than those of the logarithmic, Wald, Fieller and Bootstrap intervals.

*4.2. Negative LRs*

Tables 4 and 5 show some of the results obtained for $\omega^-$ considering the same scenarios as for $\omega^+$. Failures are indicated in bold type. From the results, the following conclusions are obtained:

  *a) Regression method*. This interval has an asymptotic behaviour which is very similar to that of the same interval for $\omega^+$.

  *b) Logarithmic CI*. In general terms, this interval can fail when $\omega^+ \neq 1$ and the dependence factors are high, whatever the sample size may be. This interval does not fail when $\omega^+ = 1$, and its *CP* is 100% or very near to this value when $n \leq 100$, and even with $n \geq 200$ if the prevalence is small. When this interval does not fail, its *AL* is lower than that of the interval obtained through regression.

  *c) Wald CI*. The Wald *CI* does not fail, and its *CP* is 100% (or very near) when $n \leq 100$, and its *CP* fluctuates around 95% when $n \geq 200$. The *AL* of the Wald *CI* is slightly lower than that of the logarithmic *CI* (when this does not fail), and its *CP* shows better fluctuations around 95% than that of the logarithmic interval.

  *c) Fieller CI*. This interval does not show any failures. In very general terms, the Fieller *CI* has a very similar *CP* to that of the Wald *CI* when $\omega^+ \neq 1$. When $\omega^+ = 1$, the *CP* of the Fieller *CI* is 100% (or near) when $n \leq 100$, and fluctuates around 95% if $n \geq 200$. Its *AL* is greater than that of the Wald *CI,* especially when $n \leq 500$.

*d) Bootstrap CI.* This interval has many failures when $\omega^+ \neq 1$, especially when the prevalence is small or moderate, and regardless of the sample size. When $\omega^+ = 1$, the interval does not fail, and its *CP* is greater than that of the Wald *CI* or the logarithmic *CI*, especially when the prevalence is small or moderate. Regarding the Fieller *CI*, the *CP* of the Bootstrap interval is very similar to that of the Fieller interval, and its *AL* is slightly lower than that of the Fieller *CI*, especially for $n \leq 500$.

*e) Bayesian CI.* The same as for $\omega^+$, the Bayesian *CI* for $\omega^-$ does not fail and has a *CP* and an *AL* which are very similar to those of the interval obtained through regression. The same as for $\omega^+$, the *CP* and the *AL* of the Bayesian interval are higher than those of the logarithmic, Wald, Fieller and Bootstrap intervals.

*4.3. Rules of application*

Considering the asymptotic behaviour of each one of the *CIs* studied, it is possible to give some general rules of application for the *CIs* studied. These rules of application are for the different scenarios considered in the simulation experiments, scenarios that correspond to realistic values of prevalence, sensitivities and specificities in clinical practice. Based on the sample size, which in practice is the only parameter set by the researcher, the rules are the following:

  a) For the ratio $\omega^+$, use the logarithmic *CI*, whatever the sample size may be, although when $n \geq 200$ we can also use the Wald, the Fieller and the Bootstrap intervals.

  b) For the ratio $\omega^-$, use the Wald *CI*, whatever the sample size may be.

Table 4. Coverage probabilities (%) and average lengths of the *CIs* for the ratio of the two negative *LRs* (I).

$LR_1^+ = 9.5$  $LR_2^+ = 4.5$  $LR_1^- = 0.056$  $LR_2^- = 0.125$  $\omega^+ = 2.111$  $\omega^- = 0.444$

$Se_1 = 0.95$  $Sp_1 = 0.90$  $Se_2 = 0.90$  $Sp_2 = 0.80$

$\pi = 10\%$  $\varepsilon_1 = 0.0225$  $\varepsilon_0 = 0.0400$

| | Regression | | Logarithmic | | Wald | | Fieller | | Bootstrap | | Bayesian | |
|---|---|---|---|---|---|---|---|---|---|---|---|---|
| n | CP | AL | CP | AL | CP | AL | CP | AL | CP | AL | CP | AL |
| 50 | 99.95 | 2.07 | 97.30 | 1.65 | 96.85 | 1.27 | 99.50 | 2.71 | **14.80** | 1.79 | 99.75 | 1.90 |
| 100 | 99.90 | 2.02 | 96.60 | 1.59 | 96.05 | 1.17 | 99.60 | 2.49 | **35.50** | 1.83 | 99.85 | 1.81 |
| 200 | 99.95 | 1.99 | 96.15 | 1.42 | 95.90 | 1.09 | 99.55 | 2.40 | **53.70** | 1.68 | 99.85 | 1.79 |
| 300 | 99.85 | 1.81 | 95.45 | 1.30 | 95.15 | 1.03 | 99.05 | 1.99 | **75.95** | 1.59 | 99.75 | 1.65 |
| 400 | 99.85 | 1.67 | 96.55 | 1.23 | 95.55 | 0.97 | 99.10 | 1.75 | **86.05** | 1.55 | 99.75 | 1.54 |
| 500 | 99.80 | 1.62 | 96.95 | 1.20 | 95.95 | 0.96 | 98.80 | 1.70 | **88.80** | 1.48 | 99.60 | 1.50 |
| 1000 | 99.55 | 1.22 | 96.90 | 0.93 | 95.90 | 0.81 | 97.85 | 1.16 | 95.80 | 1.05 | 99.45 | 1.16 |

$\pi = 10\%$  $\varepsilon_1 = 0.0360$  $\varepsilon_0 = 0.0640$

| | Regression | | Logarithmic | | Wald | | Fieller | | Bootstrap | | Bayesian | |
|---|---|---|---|---|---|---|---|---|---|---|---|---|
| n | CP | AL | CP | AL | CP | AL | CP | AL | CP | AL | CP | AL |
| 50 | 100 | 2.18 | **92.60** | 1.63 | 99.95 | 1.31 | 99.50 | 2.83 | **5.50** | 1.66 | 100 | 2.01 |
| 100 | 100 | 2.11 | **90.85** | 1.53 | 98.90 | 1.19 | 99.25 | 2.48 | **17.25** | 1.70 | 100 | 1.91 |
| 200 | 100 | 2.16 | **91.15** | 1.38 | 98.35 | 1.12 | 99.25 | 2.57 | **33.00** | 1.53 | 100 | 1.96 |
| 300 | 99.95 | 1.94 | **90.20** | 1.21 | 97.60 | 1.01 | 98.10 | 2.02 | **54.45** | 1.43 | 99.90 | 1.78 |
| 400 | 99.95 | 1.76 | **92.40** | 1.13 | 97.10 | 0.95 | 97.65 | 1.64 | **65.25** | 1.39 | 99.90 | 1.63 |
| 500 | 99.90 | 1.68 | **92.80** | 1.09 | 96.10 | 0.91 | 97.85 | 1.55 | **70.45** | 1.35 | 99.85 | 1.56 |
| 1000 | 99.90 | 1.22 | 93.40 | 0.79 | 95.60 | 0.71 | 97.45 | 0.97 | **84.65** | 0.93 | 99.80 | 1.16 |

$\pi = 25\%$  $\varepsilon_1 = 0.0225$  $\varepsilon_0 = 0.0400$

| | Regression | | Logarithmic | | Wald | | Fieller | | Bootstrap | | Bayesian | |
|---|---|---|---|---|---|---|---|---|---|---|---|---|
| n | CP | AL | CP | AL | CP | AL | CP | AL | CP | AL | CP | AL |
| 50 | 100 | 2.06 | 97.80 | 1.56 | 96.35 | 1.18 | 99.30 | 2.66 | **34.05** | 1.86 | 99.75 | 1.87 |
| 100 | 100 | 1.87 | 96.20 | 1.34 | 95.95 | 1.04 | 99.65 | 2.13 | **64.85** | 1.67 | 99.80 | 1.70 |
| 200 | 99.65 | 1.64 | 96.00 | 1.22 | 95.80 | 0.98 | 98.00 | 1.77 | **89.30** | 1.50 | 99.60 | 1.52 |
| 300 | 99.50 | 1.44 | 95.95 | 1.07 | 95.60 | 0.90 | 97.40 | 1.46 | 93.15 | 1.28 | 99.45 | 1.35 |
| 400 | 99.10 | 1.21 | 95.75 | 0.93 | 95.40 | 0.81 | 96.55 | 1.16 | 95.35 | 1.05 | 98.90 | 1.15 |
| 500 | 99.50 | 1.06 | 95.55 | 0.82 | 95.45 | 0.73 | 96.00 | 0.97 | 95.60 | 0.89 | 99.20 | 1.01 |
| 1000 | 98.60 | 0.65 | 95.20 | 0.52 | 95.15 | 0.50 | 94.65 | 0.55 | 95.55 | 0.52 | 98.45 | 0.64 |

$\pi = 25\%$  $\varepsilon_1 = 0.0360$  $\varepsilon_0 = 0.0640$

| | Regression | | Logarithmic | | Wald | | Fieller | | Bootstrap | | Bayesian | |
|---|---|---|---|---|---|---|---|---|---|---|---|---|
| n | CP | AL | CP | AL | CP | AL | CP | AL | CP | AL | CP | AL |
| 50 | 100 | 2.13 | **91.90** | 1.48 | 99.90 | 1.19 | 99.30 | 2.60 | **18.35** | 1.71 | 99.95 | 1.95 |
| 100 | 100 | 2.07 | **90.35** | 1.29 | 99.00 | 1.08 | 98.45 | 2.31 | **37.80** | 1.53 | 99.95 | 1.89 |
| 200 | 99.85 | 1.71 | **91.65** | 1.09 | 96.55 | 0.92 | 97.40 | 1.58 | **67.35** | 1.35 | 99.80 | 1.59 |
| 300 | 99.85 | 1.48 | **92.25** | 0.95 | 96.35 | 0.82 | 97.15 | 1.28 | **77.20** | 1.14 | 99.75 | 1.39 |
| 400 | 99.85 | 1.26 | **91.90** | 0.81 | 95.90 | 0.72 | 96.85 | 1.02 | **82.05** | 0.94 | 99.85 | 1.20 |
| 500 | 99.85 | 1.06 | **92.70** | 0.69 | 95.70 | 0.63 | 96.35 | 0.80 | **87.20** | 0.77 | 99.65 | 1.02 |
| 1000 | 99.50 | 0.65 | 94.45 | 0.43 | 95.35 | 0.42 | 96.20 | 0.45 | 94.40 | 0.44 | 99.55 | 0.64 |

$\pi = 50\%$  $\varepsilon_1 = 0.0225$  $\varepsilon_0 = 0.0400$

| | Regression | | Logarithmic | | Wald | | Fieller | | Bootstrap | | Bayesian | |
|---|---|---|---|---|---|---|---|---|---|---|---|---|
| n | CP | AL | CP | AL | CP | AL | CP | AL | CP | AL | CP | AL |
| 50 | 99.90 | 1.82 | 97.65 | 1.35 | 99.90 | 1.07 | 99.60 | 2.13 | **71.70** | 1.76 | 99.85 | 1.69 |
| 100 | 99.85 | 1.67 | 96.35 | 1.23 | 99.30 | 0.98 | 99.05 | 1.82 | **84.60** | 1.56 | 99.80 | 1.55 |
| 200 | 99.70 | 1.23 | 97.10 | 0.94 | 96.95 | 0.81 | 98.00 | 1.19 | 96.05 | 1.07 | 99.60 | 1.17 |
| 300 | 98.75 | 0.92 | 96.25 | 0.73 | 94.40 | 0.66 | 95.60 | 0.81 | 97.25 | 0.76 | 98.50 | 0.89 |
| 400 | 98.55 | 0.75 | 95.45 | 0.60 | 94.45 | 0.56 | 95.25 | 0.64 | 96.80 | 0.61 | 98.60 | 0.73 |
| 500 | 98.15 | 0.66 | 94.35 | 0.53 | 94.40 | 0.50 | 94.10 | 0.55 | 95.05 | 0.53 | 97.80 | 0.65 |
| 1000 | 98.65 | 0.44 | 95.20 | 0.35 | 95.20 | 0.35 | 94.80 | 0.36 | 94.35 | 0.36 | 98.45 | 0.43 |

$\pi = 50\%$  $\varepsilon_1 = 0.0360$  $\varepsilon_0 = 0.0640$

| | Regression | | Logarithmic | | Wald | | Fieller | | Bootstrap | | Bayesian | |
|---|---|---|---|---|---|---|---|---|---|---|---|---|
| n | CP | AL | CP | AL | CP | AL | CP | AL | CP | AL | CP | AL |
| 50 | 100 | 1.90 | **92.35** | 1.25 | 99.30 | 1.04 | 98.35 | 2.01 | **47.90** | 1.60 | 99.95 | 1.77 |
| 100 | 100 | 1.74 | **92.05** | 1.11 | 97.80 | 0.93 | 97.80 | 1.63 | **60.20** | 1.43 | 99.95 | 1.62 |
| 200 | 100 | 1.26 | 93.55 | 0.82 | 96.30 | 0.73 | 97.45 | 1.02 | **81.85** | 0.97 | 99.90 | 1.20 |
| 300 | 99.65 | 0.94 | 94.70 | 0.62 | 95.15 | 0.58 | 96.65 | 0.70 | **90.15** | 0.67 | 99.50 | 0.91 |
| 400 | 99.70 | 0.77 | 94.55 | 0.51 | 95.30 | 0.48 | 95.95 | 0.54 | 93.10 | 0.52 | 99.50 | 0.75 |
| 500 | 99.75 | 0.65 | 95.30 | 0.44 | 95.20 | 0.42 | 95.85 | 0.46 | 94.80 | 0.44 | 99.55 | 0.64 |
| 1000 | 99.65 | 0.43 | 95.75 | 0.30 | 94.80 | 0.29 | 95.40 | 0.30 | 96.30 | 0.29 | 99.55 | 0.43 |

Table 5. Coverage probabilities (%) and average lengths of the *CIs* for the ratio of the two negative *LRs* (II).

$LR_1^+ = 6$  $LR_2^+ = 6$  $LR_1^- = 0.118$  $LR_2^- = 0.118$  $\omega^+ = 1$  $\omega^- = 1$

$Se_1 = 0.90$  $Sp_1 = 0.85$  $Se_2 = 0.90$  $Sp_2 = 0.85$

$\pi = 10\%$  $\varepsilon_1 = 0.0450$  $\varepsilon_0 = 0.0638$

| | Regression | | Logarithmic | | Wald | | Fieller | | Bootstrap | | Bayesian | |
|---|---|---|---|---|---|---|---|---|---|---|---|---|
| n | CP | AL | CP | AL | CP | AL | CP | AL | CP | AL | CP | AL |
| 50 | 100 | 2.55 | 100 | 1.84 | 99.50 | 1.55 | 100 | 3.35 | 100 | 1.62 | 100 | 2.34 |
| 100 | 100 | 2.54 | 100 | 1.74 | 98.85 | 1.44 | 99.95 | 3.04 | 100 | 1.65 | 100 | 2.33 |
| 200 | 100 | 2.57 | 100 | 1.58 | 95.90 | 1.36 | 99.90 | 3.01 | 100 | 1.56 | 100 | 2.36 |
| 300 | 100 | 2.48 | 100 | 1.52 | 93.85 | 1.34 | 99.60 | 2.70 | 100 | 1.51 | 100 | 2.31 |
| 400 | 100 | 2.39 | 99.65 | 1.51 | 93.15 | 1.34 | 99.20 | 2.53 | 100 | 1.51 | 99.90 | 2.25 |
| 500 | 100 | 2.39 | 99.65 | 1.53 | 94.35 | 1.37 | 99.05 | 2.55 | 100 | 1.51 | 100 | 2.26 |
| 1000 | 99.85 | 1.98 | 97.15 | 1.33 | 93.95 | 1.24 | 96.85 | 1.86 | 98.70 | 1.38 | 99.85 | 1.91 |

$\pi = 10\%$  $\varepsilon_1 = 0.0720$  $\varepsilon_0 = 0.1020$

| | Regression | | Logarithmic | | Wald | | Fieller | | Bootstrap | | Bayesian | |
|---|---|---|---|---|---|---|---|---|---|---|---|---|
| n | CP | AL | CP | AL | CP | AL | CP | AL | CP | AL | CP | AL |
| 50 | 100 | 2.73 | 100 | 1.76 | 99.90 | 1.56 | 100 | 3.45 | 100 | 1.39 | 100 | 2.51 |
| 100 | 100 | 2.68 | 100 | 1.56 | 99.80 | 1.40 | 100 | 2.84 | 100 | 1.34 | 100 | 2.49 |
| 200 | 100 | 2.80 | 100 | 1.42 | 99.80 | 1.30 | 100 | 2.98 | 100 | 1.23 | 100 | 2.59 |
| 300 | 100 | 2.71 | 100 | 1.28 | 98.60 | 1.19 | 99.95 | 2.62 | 100 | 1.12 | 100 | 2.53 |
| 400 | 100 | 2.52 | 100 | 1.19 | 97.70 | 1.11 | 98.90 | 2.05 | 100 | 1.07 | 100 | 2.37 |
| 500 | 100 | 2.42 | 100 | 1.13 | 97.10 | 1.07 | 97.95 | 1.85 | 100 | 1.03 | 100 | 2.29 |
| 1000 | 100 | 1.91 | 99.80 | 0.85 | 96.80 | 0.82 | 97.15 | 1.16 | 100 | 0.80 | 100 | 1.85 |

$\pi = 25\%$  $\varepsilon_1 = 0.0450$  $\varepsilon_0 = 0.0638$

| | Regression | | Logarithmic | | Wald | | Fieller | | Bootstrap | | Bayesian | |
|---|---|---|---|---|---|---|---|---|---|---|---|---|
| n | CP | AL | CP | AL | CP | AL | CP | AL | CP | AL | CP | AL |
| 50 | 100 | 2.56 | 100 | 1.72 | 98.20 | 1.46 | 100 | 3.23 | 100 | 1.67 | 100 | 2.36 |
| 100 | 100 | 2.57 | 100 | 1.53 | 95.45 | 1.35 | 99.85 | 2.91 | 100 | 1.52 | 99.95 | 2.38 |
| 200 | 99.95 | 2.40 | 99.50 | 1.54 | 93.90 | 1.38 | 98.90 | 2.57 | 99.95 | 1.53 | 99.90 | 2.26 |
| 300 | 99.85 | 2.26 | 98.55 | 1.48 | 94.65 | 1.35 | 98.00 | 2.35 | 99.75 | 1.47 | 99.80 | 2.15 |
| 400 | 99.70 | 1.98 | 96.95 | 1.33 | 93.05 | 1.24 | 96.20 | 1.85 | 98.20 | 1.37 | 99.55 | 1.92 |
| 500 | 99.55 | 1.74 | 95.20 | 1.18 | 92.35 | 1.11 | 94.55 | 1.50 | 96.40 | 1.24 | 99.40 | 1.70 |
| 1000 | 99.25 | 1.15 | 94.80 | 0.79 | 94.25 | 0.77 | 94.40 | 0.86 | 94.15 | 0.84 | 99.20 | 1.13 |

$\pi = 25\%$  $\varepsilon_1 = 0.0720$  $\varepsilon_0 = 0.1020$

| | Regression | | Logarithmic | | Wald | | Fieller | | Bootstrap | | Bayesian | |
|---|---|---|---|---|---|---|---|---|---|---|---|---|
| n | CP | AL | CP | AL | CP | AL | CP | AL | CP | AL | CP | AL |
| 50 | 100 | 2.76 | 100 | 1.57 | 99.80 | 1.41 | 100 | 3.11 | 100 | 1.40 | 100 | 2.56 |
| 100 | 100 | 2.81 | 100 | 1.37 | 98.90 | 1.26 | 100 | 2.95 | 100 | 1.22 | 100 | 2.61 |
| 200 | 100 | 2.45 | 100 | 1.14 | 97.75 | 1.07 | 100 | 1.86 | 100 | 1.04 | 100 | 2.32 |
| 300 | 100 | 2.22 | 99.90 | 1.01 | 97.20 | 0.97 | 99.80 | 1.54 | 100 | 0.93 | 100 | 2.13 |
| 400 | 100 | 1.92 | 96.95 | 0.86 | 96.80 | 0.83 | 98.55 | 1.17 | 99.95 | 0.80 | 100 | 1.86 |
| 500 | 100 | 1.69 | 96.55 | 0.74 | 96.45 | 0.72 | 98.15 | 0.94 | 99.90 | 0.71 | 100 | 1.65 |
| 1000 | 100 | 1.13 | 96.05 | 0.49 | 95.95 | 0.48 | 96.50 | 0.53 | 98.45 | 0.49 | 100 | 1.12 |

$\pi = 50\%$  $\varepsilon_1 = 0.0450$  $\varepsilon_0 = 0.0638$

| | Regression | | Logarithmic | | Wald | | Fieller | | Bootstrap | | Bayesian | |
|---|---|---|---|---|---|---|---|---|---|---|---|---|
| n | CP | AL | CP | AL | CP | AL | CP | AL | CP | AL | CP | AL |
| 50 | 100 | 2.43 | 100 | 1.59 | 95.50 | 1.39 | 99.90 | 2.80 | 100 | 1.65 | 99.95 | 2.29 |
| 100 | 99.95 | 2.44 | 99.25 | 1.56 | 94.70 | 1.40 | 99.00 | 2.69 | 99.95 | 1.55 | 99.90 | 2.31 |
| 200 | 99.80 | 2.01 | 96.90 | 1.34 | 93.30 | 1.25 | 96.20 | 1.89 | 98.85 | 1.37 | 99.70 | 1.94 |
| 300 | 99.65 | 1.57 | 96.75 | 1.08 | 94.30 | 1.03 | 96.30 | 1.29 | 97.20 | 1.15 | 99.60 | 1.54 |
| 400 | 99.70 | 1.32 | 95.40 | 0.91 | 94.65 | 0.88 | 95.20 | 1.02 | 95.20 | 0.97 | 99.70 | 1.30 |
| 500 | 99.70 | 1.17 | 95.10 | 0.81 | 94.90 | 0.78 | 94.70 | 0.88 | 94.20 | 0.85 | 99.65 | 1.15 |
| 1000 | 99.40 | 0.78 | 95.20 | 0.54 | 94.60 | 0.54 | 95.05 | 0.56 | 94.75 | 0.56 | 99.35 | 0.77 |

$\pi = 50\%$  $\varepsilon_1 = 0.0720$  $\varepsilon_0 = 0.1020$

| | Regression | | Logarithmic | | Wald | | Fieller | | Bootstrap | | Bayesian | |
|---|---|---|---|---|---|---|---|---|---|---|---|---|
| n | CP | AL | CP | AL | CP | AL | CP | AL | CP | AL | CP | AL |
| 50 | 100 | 2.67 | 99.95 | 1.36 | 99.50 | 1.26 | 99.95 | 2.64 | 100 | 1.30 | 100 | 2.51 |
| 100 | 100 | 2.49 | 100 | 1.16 | 98.45 | 1.09 | 100 | 1.95 | 100 | 1.09 | 100 | 2.36 |
| 200 | 100 | 1.94 | 99.55 | 0.86 | 97.30 | 0.83 | 99.40 | 1.18 | 100 | 0.81 | 100 | 1.88 |
| 300 | 100 | 1.55 | 98.80 | 0.67 | 97.00 | 0.66 | 98.55 | 0.80 | 99.75 | 0.65 | 100 | 1.51 |
| 400 | 100 | 1.30 | 96.95 | 0.56 | 96.90 | 0.55 | 97.80 | 0.63 | 99.60 | 0.55 | 100 | 1.28 |
| 500 | 100 | 1.14 | 96.25 | 0.50 | 96.25 | 0.49 | 96.05 | 0.54 | 98.20 | 0.50 | 100 | 1.13 |
| 1000 | 100 | 0.78 | 95.35 | 0.34 | 95.10 | 0.34 | 95.35 | 0.35 | 95.30 | 0.35 | 100 | 0.77 |

## 5. Sample size

An important question when comparing two parameters is the calculation of the sample size necessary to compare the parameters with a determined error and power. In the context of the comparison of the *LRs*, Roldán-Nofuentes and Luna (2007) proposed a method to calculate the sample size to solve the hypothesis test $H_0 : \ln(\omega) = 0$ vs $H_1 : \ln(\omega) \neq 0$. We then study the same problem but from the perspective of the *CIs*. Therefore, we study the problem of calculating the sample size necessary to estimate the ratio between the two *LRs* with a precision $\delta$ and a confidence $100(1-\alpha)\%$. As in the previous sections, we consider that $\omega$ is $\omega^+$ or $\omega^-$. Let us first consider the Wald *CI*, which can be applied both to estimate $\omega^+$ (with $n \geq 200$) and $\omega^-$ (for any sample size). Based on the asymptotic normality of the estimator of $\omega$, it is verified that $\hat{\omega} \in \omega \pm z_{1-\alpha/2}\sqrt{Var(\hat{\omega})}$, i.e. the probability of obtaining an estimator $\hat{\omega}$ is in this interval with a probability $100(1-\alpha)\%$. Let us consider that $LR_2 > LR_1$ and, therefore, that $\omega < 1$ (the Wald interval will be lower than one) and let $\delta$ be the precision set by the researcher. As it has been assumed that $\omega < 1$, then $\delta$ must be lower than one, and if we want to have a high level of precision then $\delta$ must be a small value. The sample size *n* is calculated from the expression

$$\delta = z_{1-\alpha/2}\omega\sqrt{\frac{Var(\hat{LR}_1)}{LR_1^2} + \frac{Var(\hat{LR}_2)}{LR_2^2} - \frac{2Cov(\hat{LR}_1, \hat{LR}_2)}{LR_1 LR_2}}. \tag{21}$$

This equation is obtained from the Wald *CI* (equation (12)). Substituting the variances and the covariance with their respective expressions given in equations (6) and clearing $n$ we obtain the expression of the sample size to estimate $\omega$ with a precision $\delta$ and a confidence $100(1-\alpha)\%$. For $\omega^+$ the equation of the sample size is

$$n = \left(\frac{z_{1-\alpha/2}\omega^+}{\delta}\right)^2 \left[\sum_{h=1}^{2}\left(\frac{1-Se_h}{\pi Se_h} + \frac{Sp_h}{\bar{\pi}(1-Sp_h)}\right) - \frac{2\varepsilon_1}{\pi Se_1 Se_2} - \frac{2\varepsilon_0}{\bar{\pi}(1-Sp_1)(1-Sp_2)}\right], \quad (22)$$

and for $\omega^-$ is

$$n = \left(\frac{z_{1-\alpha/2}\omega^-}{\delta}\right)^2 \left[\sum_{h=1}^{2}\left(\frac{Se_h}{\pi(1-Se_h)} + \frac{1-Sp_h}{\bar{\pi}Sp_h}\right) - \frac{2\varepsilon_1}{\pi(1-Se_1)(1-Se_2)} - \frac{2\varepsilon_0}{\bar{\pi}Sp_1 Sp_2}\right]. \quad (23)$$

If it is considered that $\omega > 1$ (and consequently the Wald CI is higher than one) the BDTs can always be permuted and $\omega$ will then be lower than one. Another alternative consists of setting a value for a precision $\delta'$, in a similar way to the previous situation when $\omega < 1$, and then apply equation (22) or (23) considering $\delta = \hat{\omega}^2 \delta'$. As is explained at the end of Section 3, this is due to the fact that if $(L_\omega, U_\omega)$ is the Wald CI for $\omega = LR_1/LR_2 < 1$ then the Wald CI for $\omega' = 1/\omega = LR_2/LR_1$ is $\left(\frac{L_\omega}{\hat{\omega}^2}, \frac{U_\omega}{\hat{\omega}^2}\right)$. It is easy to check that the calculated value of the sample size $n$ is the same both if $\omega < 1$ (with a precision $\delta$) and if $\omega > 1$ (with precision $\delta = \hat{\omega}^2 \delta'$).

In order to be able to apply the previous equations, it is necessary to know the sensitivities, the specificities (and therefore the *LRs*, $\omega^+$ and $\omega^-$), the dependence factors between the two *BDTs* $(\varepsilon_i)$ and the prevalence $(\pi)$. In practice, these values can be estimated from a pilot sample or can be obtained from another similar study. Therefore, the method to calculate the sample size requires us to know some estimations of the accuracy (*Se* and *Sp*) of each *BDT*, of the dependence factors between the *BDTs* and of the disease prevalence, obtained for example from a pilot study or from other previous studies. The method to calculate the size of the sample consists of the following steps:

Step 1. Take a pilot sample sized $n_0$ (in general terms, $n_0 \geq 200$ if $\omega^+$ is estimated to then be able to calculate the Wald *CI*), and with this sample we calculate $\hat{Se}_h$, $\hat{Sp}_h$ (and therefore $\hat{LR}_h$, $\hat{\omega}^+$ and $\hat{\omega}^-$), $\hat{\varepsilon}_i$ and $\hat{\pi}$. The Wald *CI* for $\omega$ is then calculated, and if this interval has a precision $\delta$, i.e. $z_{1-\alpha/2}\sqrt{\hat{Var}(\hat{\omega})} \leq \delta$, then the required precision has been reached; if not, go to the following step.

Step 2. Based on the estimations obtained in Step 1, calculate the sample size $n$ applying equation (22) or (23).

Step 3. Take the sample of $n$ individuals (add $n - n_0$ individuals to the initial pilot sample), and from this new sample we calculate $\hat{Se}_h$, $\hat{Sp}_h$, $\hat{\varepsilon}_i$, $\hat{\pi}$ and the Wald *CI*. If the Wald *CI* has a precision $\delta$, then the set precision has been achieved; if not, consider the new sample to be a pilot sample $(n_0 = n)$ and go back to Step 1.

This proposed procedure to calculate the sample size is iterative, and therefore it does not guarantee that with the sample size calculated we can then estimate the parameter $\omega$ with the required precision. Moreover, if the researcher sets a precision $\delta^+$ to estimate $\omega^+$ and also sets a precision $\delta^-$ to estimate $\omega^-$, once both sample sizes have been calculated through the previous method, the researcher must take a sample size of at least the maximum of the two sample sizes, to thus guarantee the precision in both estimations. In general, the calculation of the sample size makes sense when the confidence interval for $\omega$ does not contain the value one, since in this situation (the interval contains the value one) the equality of both *LRs* is not rejected and it does not make sense to determine how much larger one *LR* is compared to the other. Nevertheless, if the pilot sample is small (for example to estimate $\omega^-$) and the Wald *CI* for $\omega^-$ contains the value 1, it may be useful to calculate the sample size to estimate the $\omega^-$. In this situation, the Wald *CI* for $\omega^-$ will be very wide (as the pilot sample is small) and may contain the value 1 even if $LR_1^-$ and $LR_2^-$ are different.

The calculation of the sample size depends on the estimations obtained from an initial pilot sample. In order to study the effect that this sample has on the calculation of the sample size, simulation experiments were carried out which were similar to those carried out in Section 4. From the values of the parameters, we calculated the sample size *n* applying equation (22) or (23) depending on the case, taking a precision equal to 0.10, and we then generated $N = 10,000$ random samples with multinomial distributions sized *n*. In each one of the *N* random samples, we calculated the sample size $n_i'$ from the estimators calculated with the random sample, and then calculated the average sample size $\bar{n} = \sum n_i'/N$ and the relative bias $RB(n') = (\bar{n} - n)/n$. Table 6 shows the results obtained for the scenarios considered in Tables 5 and 6 ($\omega \neq 1$). From the results, it holds that that the dependence factors $\varepsilon_i$ have an important effect on the calculation of the sample size, and the sample size is smaller when the dependence factors are larger. Moreover, the increase in the prevalence means an increase (decrease) in the sample size to estimate $\omega^+$ ($\omega^-$). The relative biases obtained are very small, and therefore the sample sizes calculated from equations (22) and (23) are robust. Consequently, the initial pilot sample does not have an important effect on the determination of the sample size to estimate $\omega$.

Table 6. Sample size to estimate $\omega$.

| $LR_1^+ = 9.5$ | $LR_2^+ = 4.5$ | $LR_1^- = 0.056$ | $LR_2^- = 0.125$ | $\omega^+ = 2.111$ | $\omega^- = 0.444$ |
|---|---|---|---|---|---|
| $Se_1 = 0.95$ | $Sp_1 = 0.90$ | $Se_2 = 0.90$ | $Sp_2 = 0.80$ | | |

| Sample size for $\omega^+$ | | | |
|---|---|---|---|
| $\varepsilon_1 = 0.0225$ $\varepsilon_0 = 0.0400$ | | | |
| | $\pi = 10\%$ | $\pi = 25\%$ | $\pi = 50\%$ |
| Sample size | 958 | 1,073 | 1,571 |
| Average sample size | 981 | 1,084 | 1,597 |
| Relative bias (%) | 2.40 | 1.03 | 1.66 |
| $\varepsilon_1 = 0.0360$ $\varepsilon_0 = 0.0640$ | | | |
| | $\pi = 10\%$ | $\pi = 25\%$ | $\pi = 50\%$ |
| Sample size | 701 | 786 | 1,152 |
| Average sample size | 734 | 796 | 1,160 |
| Relative bias (%) | 4.71 | 1.27 | 0.69 |
| Sample size for $\omega^-$ | | | |
| $\varepsilon_1 = 0.0225$ $\varepsilon_0 = 0.0400$ | | | |
| | $\pi = 10\%$ | $\pi = 25\%$ | $\pi = 50\%$ |
| Sample size | 14,439 | 5,793 | 2,922 |
| Average sample size | 14,715 | 5,896 | 2,966 |
| Relative bias (%) | 1.91 | 1.78 | 1.51 |
| $\varepsilon_1 = 0.0360$ $\varepsilon_0 = 0.0640$ | | | |
| | $\pi = 10\%$ | $\pi = 25\%$ | $\pi = 50\%$ |
| Sample size | 10,336 | 4,147 | 2,092 |
| Average sample size | 10,482 | 4,186 | 2,118 |
| Relative bias (%) | 1.41 | 0.94 | 1.24 |

If the initial pilot sample has a small or moderate size, then in order to estimate $\omega^+$ we use the logarithmic *CI*. In this situation, the process is similar to the previous one, and the sample size is calculated from the equation $\ln(\delta) = z_{1-\alpha/2}\sqrt{Var\left[\ln(\hat{\omega}^+)\right]}$, where the expression of $Var\left[\ln(\hat{\omega}^+)\right]$ is given in equation (10). Following a similar process to the previous one, it holds that

$$n = \left(\frac{z_{1-\alpha/2}}{\ln(\delta)}\right)^2 \left[\sum_{h=1}^{2}\left(\frac{1-Se_h}{\pi Se_h} + \frac{Sp_h}{\bar{\pi}(1-Sp_h)}\right) - \frac{2\varepsilon_1}{\pi Se_1 Se_2} - \frac{2\varepsilon_0}{\bar{\pi}(1-Sp_1)(1-Sp_2)}\right]. \quad (24)$$

# 6. Applications

The results obtained were applied to two real examples: a) a study of the diagnosis of coronary disease, and another study of the diagnosis of colorectal cancer.

*6.1. Diagnosis of coronary disease*

The results obtained were applied to the study by Weiner et al (1979) on the diagnosis of coronary disease, which is a widely used study to illustrate statistical methods for the estimation and comparison of parameters of *BDTs*. Weiner et al studied the diagnosis of coronary artery disease using as diagnostic tests the exercise test and the resting *EKG*, and the coronary arteriography as a *GS*. Table 7 shows the frequencies obtained by applying three medical tests to a sample of 1,465 males, where $T_1$ models the result of the exercise test, $T_2$ models the result of the resting *EKG* and *D* the result of the *GS*. Table 7 also shows the estimations of the *LRs* $(\omega)$ and their standard errors, as well as the *CIs* for $\omega^+$ and $\omega^-$.

For $\omega^+$, from any of the six *CIs* (all of them are greater than one) it holds that the positive *LR* of the exercise test is significantly larger than the positive *LR* of the resting *EKG*, i.e. a positive result in the exercise test is more indicative of the presence of the disease than a positive result in the resting *EKG*. Interpreting the results of the logarithmic *CI* it holds that the positive *LR* of the exercise test is (with a confidence of 95%) a value between 1.632 and 2.713 times larger than the positive *LR* of the resting *EKG*.

Regarding $\omega^-$, all of the *CIs* intervals (all are less than one) we reject the equality of the two negative *LRs*, and it holds that a negative result for the resting *EKG* is more indicative of the absence of the disease than a negative result of the exercise test. Interpreting the Wald *CI*, the negative *LR* of the resting *EKG* is (with a confidence of 95%) a value between 2.872 $\left(=0.262/0.302^2\right)$ and 3.783 $\left(=0.345/0.302^2\right)$ times larger than the negative *LR* of the exercise test.

Moreover, in order to illustrate the method to calculate the sample size, we are going to consider that the researcher wants to estimate $\omega^+$ with a precision equal to 0.10, which can be considered to be a high precision. The Wald *CI* for $\omega^+$ is $(1.569, 2.639)$, and therefore multiplying this interval by $1/\left(\hat{\omega}^+\right)^2 = 1/2.109^2$ it holds that the 95% Wald *CI* for $\omega'^+ = LR_2^+/LR_1^+$ is $(0.353, 0.593)$, and the precision is 0.12. As 0.12 is higher than 0.10, it is necessary to increase the sample size to estimate $\omega^+$ with the required precision. Setting the confidence at 95% and taking $\delta = \left(\hat{\omega}^+\right)^2 \delta' = 2.109^2 \times 0.10 \approx 0.445$, applying equation (22) it holds that $n = 2{,}146$. Consequently, it is necessary to add 681 new individuals to the initial sample of 1,465 individuals, and once the data are obtained it is necessary to check that the required precision has been achieved.

Table 7. Diagnosis of coronary disease.

|  | $T_1 = 1$ | | $T_1 = 0$ | | |
|---|---|---|---|---|---|
|  | $T_2 = 1$ | $T_2 = 0$ | $T_2 = 1$ | $T_2 = 0$ | Total |
| $D = 1$ | 224 | 591 | 32 | 176 | 1,023 |
| $D = 0$ | 35 | 80 | 41 | 286 | 442 |
| Total | 259 | 671 | 73 | 462 | 1,465 |

| Results | | | | |
|---|---|---|---|---|
|  | $Se$ | $Sp$ | $LR^+$ | $LR^-$ |
| *Exercise test* | $0.797 \pm 0.013$ | $0.740 \pm 0.021$ | $3.065 \pm 0.250$ | $0.274 \pm 0.019$ |
| *Resting EKG* | $0.250 \pm 0.014$ | $0.828 \pm 0.018$ | $1.453 \pm 0.171$ | $0.906 \pm 0.026$ |
| $p$ | $\varepsilon_1$ | $\varepsilon_0$ | $\omega^+ = LR_1^+/LR_2^+$ | $\omega^- = LR_1^-/LR_2^-$ |
| 0.698 | 0.020 | 0.034 | $2.109 \pm 0.273$ | $0.302 \pm 0.021$ |

| CIs for $\omega^+ = LR_1^+/LR_2^+$ | | |
|---|---|---|
| Regression *CI* | Logarithmic *CI* | Wald *CI* |
| $(1.589 , 2.786)$ | $(1.632 , 2.713)$ | $(1.569 , 2.639)$ |
| Fieller *CI* | Bootstrap *CI* | Bayesian *CI* |
| $(1.647 , 2.765)$ | $(1.501 , 2.612)$ | $(1.668 , 2.567)$ |

| CIs for $\omega^- = LR_1^-/LR_2^-$ | | |
|---|---|---|
| Regression *CI* | Logarithmic *CI* | Wald *CI* |
| $(0.263 , 0.351)$ | $(0.265 , 0.348)$ | $(0.262 , 0.345)$ |
| Fieller *CI* | Bootstrap *CI* | Bayesian *CI* |
| $(0.262 , 0.346)$ | $(0.280 , 0.348)$ | $(0.264 , 0.343)$ |

*6.2. Diagnosis of colorectal cancer*

The results obtained were applied to a study of the diagnosis of colorectal cancer, using as diagnostic tests *Fecal Occult Blood Testing* (*FOBT*) and *Fecal Immunochemical Testing* (*FIT*), and the biopsy as the *GS*. Table 8 shows the results obtained by applying the three tests to a sample of 168 adult men with suspicious symptoms of the disease, where the variable $T_1$ models the result of the *FOBT*, $T_2$ models the result of the *FIT* and *D* models the result of the biopsy. This data came from a study carried out at the University Hospital of Granada in Spain. Table 8 also shows the estimations of the *LRs*, their standard errors and the confidence intervals for $\omega^+$ and $\omega^-$.

Applying the rule given in Section 4.3, as $n = 168 < 200$ the logarithmic *CI* for $\omega^+$ must be used in addition to the Wald *CI* for $\omega^-$. For $\omega^+$, the logarithmic *CI* contains the value one, and therefore we do not reject the equality of both positive *LRs*. Regarding $\omega^-$, the Wald *CI* does not contain the value one, and therefore we reject the equality of both negative *LRs*. Thus, a negative result for the *FOBT* is more indicative of the presence of colorectal cancer than a negative result for the *FIT*. The negative *LR* of the *FOBT* is (with a confidence of 95%) a value between 1.321 and 3.183 times larger than the negative *LR* of the *FIT*. The Wald *CI* for $1/\omega^-$ is $(0.260, 0.628)$, calculated as $(1.321/2.252^2, 3.183/2.252^2)$.

In order to illustrate in this example the method of sample size calculation, let us suppose that the researchers want to estimate $1/\omega^-$ with a precision equal to 0.10, or in other words, to estimate $\omega^-$ with a precision of $0.10 \times (\hat{\omega}^-)^2 = 0.10 \times 2.252^2 \approx 0.50$. As with the sample of 168 individuals the precision obtained with the Wald *CI* for $\omega^-$ is $0.931 > 0.50$, or rather a precision equal to 0.184 $(> 0.10)$ with the Wald *CI* for $1/\omega^-$, then it is necessary to calculate the sample size. Considering the sample of 168 individuals to be a pilot sample, applying equation (23) it holds that $n = 561$. Therefore, 561 individuals are needed (we have to add 393 to the sample of 198) in order to estimate $\omega^-$ $(1/\omega^-)$ with a precision equal to 0.50 (0.10) with a confidence of 95%.

Table 8. Diagnosis of colorectal cancer.

| | $T_1 = 1$ | | $T_1 = 0$ | | |
|---|---|---|---|---|---|
| | $T_2 = 1$ | $T_2 = 0$ | $T_2 = 1$ | $T_2 = 0$ | Total |
| $D = 1$ | 68 | 1 | 18 | 13 | 100 |
| $D = 0$ | 4 | 2 | 1 | 61 | 68 |
| Total | 72 | 3 | 19 | 74 | 168 |

| Results | | | | |
|---|---|---|---|---|
| | $Se$ | $Sp$ | $LR^+$ | $LR^-$ |
| $FOBT$ | $0.690 \pm 0.046$ | $0.912 \pm 0.034$ | $7.841 \pm 3.093$ | $0.340 \pm 0.052$ |
| $FIT$ | $0.860 \pm 0.035$ | $0.926 \pm 0.032$ | $11.622 \pm 5.057$ | $0.151 \pm 0.038$ |
| $p$ | $\varepsilon_1$ | $\varepsilon_0$ | $\omega^+ = LR_1^+ / LR_2^+$ | $\omega^- = LR_1^- / LR_2^-$ |
| 0.595 | 0.087 | 0.052 | $0.675 \pm 0.215$ | $2.252 \pm 0.475$ |

| CIs for $\omega^+ = LR_1^+ / LR_2^+$ | | |
|---|---|---|
| Regression $CI$ | Logarithmic $CI$ | Wald $CI$ |
| $(0.212, 2.108)$ | $(0.356, 1.255)$ | $(0.254, 1.096)$ |
| Fieller $CI$ | Bootstrap $CI$ | Bayesian $CI$ |
| $(0.278, 2.277)$ | $(0.281, 1.283)$ | $(0.222, 2.057)$ |

| CIs for $\omega^- = LR_1^- / LR_2^-$ | | |
|---|---|---|
| Regression $CI$ | Logarithmic $CI$ | Wald $CI$ |
| $(1.265, 4.001)$ | $(1.488, 3.403)$ | $(1.321, 3.183)$ |
| Fieller $CI$ | Bootstrap $CI$ | Bayesian $CI$ |
| $(1.556, 3.894)$ | $(1.553, 3.778)$ | $(1.281, 4.006)$ |

## 7. Discussion

The *LRs* are parameters that are used to assess and compare the effectiveness of *BDTs*, and only depend on the accuracy (sensitivity and specificity) of the *BDT*. The comparison of the positive (negative) *LRs* of two *BDTs* subject to a paired design is a topic which has not been widely studied in Statistical literature and consists of the comparison of two relative risks subject to the same type of design. The previous studies (Leisenring and Pepe (1998) and Pepe (2003), Roldán-Nofuentes and Luna (2007), Dolgun et al (2012) focused mainly on the study of hypothesis tests to compare the positive (negative) *LRs* of the two *BDTs*. The comparison of the positive (negative) *LRs* through *CIs* has been the object of the very little research, and the studies that have been published by Pepe (2003) and Roldán-Nofuentes and Luna (2007) have focused on proposing *CIs* without dealing with this question in more depth. In this article, we extend the scope of these previous studies, proposing four new intervals: three of which are frequentist (Wald, Fieller and Bootstrap) and one which is Bayesian. The Wald and Fieller intervals are based on the asymptotic normality of the ratio of the *LRs*, and the Bootstrap interval is based on the fact that the bootstrap estimator of the ratio of the *LRs* can be transformed to a normal distribution. Regarding the Bayesian Interval, this was obtained by applying the Monte Carlo method considering a priori non-informative distributions. The importance of the study of the *CIs* for the ratio of the positive (negative) *LRs* does not only lie in the fact that these *CIs* allow us to compare the two positive (negative) *LRs*, but also that it allows us to determine (when the equality of both *LRs* is rejected) how much bigger one *LR* than the other, which means an advantage over the hypothesis tests.

The comparison of the asymptotic behaviour of the six *CIs* was studied through simulation experiments. The results of these experiments has shown that, in th scenarios considered, in order to estimate the ratio $\omega^+ = LR_1^+ / LR_2^+$, in general terms, the intervals with the best behaviour are the logarithmic one (for all the sample sizes), the Wald, Fieller or Bootstrap intervals (these last three for large or very large samples); whereas in order to estimate $\omega^- = LR_1^- / LR_2^-$ the interval with the best behaviour is the Wald interval (for all of the samples sizes). The use of different *CIs* for $\omega^+$ and for $\omega^-$ may be due to the convergence to the normal distribution of the estimators. For an informative *BDT*, i.e. for a *BDT* whose Youden index is higher than 0 $(Y = Se + Sp - 1 > 0)$, it must be verified that $LR^+ > 1$ and that $LR^- < 1$. Then, considering that the two *BDTs* are informative (as should be the case in clinical practice), $\omega^+$ is the ratio between two values greater than 1 and $\omega^-$ is the ratio between two values lower than 1. For $\omega^+$, $\ln \hat{\omega}^+$ converges better to the normal distribution than $\hat{\omega}^+$ for $n<200$, but when $n \geq 200$ both ($\hat{\omega}^+$ and $\ln\hat{\omega}^+$) has a good approximation to the normal distribution. The Wald *CI* for $\omega^-$ has a better asymptotic behaviour than the logarithmic *CI* for $\omega^-$, which must be due to the fact that $\hat{\omega}^-$ converges more quickly to the normal distribution (even with large samples) than $\ln\hat{\omega}^-$.

An important question when comparing parameters of two *BDTs* is the calculation of the sample size necessary to compare the parameters based on certain specifications. When a hypothesis test is carried out, the sample size is calculated based on an error $\alpha$, a power $\theta$ and a difference (or ratio) to be detected among the parameters. Roldán-Nofuentes and Luna (2007) proposed a method to calculate sample size to solve the hypothesis test ($H_0: \ln\omega = 0$) of equality of the positive (negative) *LRs*. This article proposes, as a complement to the study of the *CIs*, a method to determine the sample size necessary to estimate the ratio between the *LRs* with a previously set precision. This is a topic that has never been studied and, therefore, represents a contribution to Statistical literature on the subject analysed in this article. The method, which is based on the Wald (logarithmic) *CI*, requires knowledge of the estimations of the sensitivities, specificities, dependence factors and disease prevalence. These estimations can be obtained from a pilot sample or another similar study and, therefore, as it depends on the pilot sample selected, the method does not guarantee that the sample size calculated will be estimated with the set precision, and it is necessary to check the precision.

The intervals studied in this article can also be applied when the sample design is case-control. In this type of design, the two *BDTs* are applied to all of the individuals in two random samples, one of $n_1$ individuals with the disease and another one of $n_2$ individuals without the disease. If thus type of sampling is used, two multinomial distributions are involved, one for the case sample, whose probabilities are

$$p_{ij} = Se_1^i (1-Se_1)^{1-i} Se_2^j (1-Se_2)^{1-j} + \delta_{ij}\varepsilon_1$$ con $\sum p_{ij} = 1$, and the other for the control sample, whose probabilities are $q_{ij} = Sp_1^{1-i}(1-Sp_1)^i Sp_2^{1-j}(1-Sp_2)^j + \delta_{ij}\varepsilon_0$ with $\sum q_{ij} = 1$. Here, the variances-covariances of the sensitivities and specificities are

$$Var(\hat{Se}_h) = \frac{Se_h(1-Se_h)}{n_1}, \quad Var(\hat{Sp}_h) = \frac{Sp_h(1-Sp_h)}{n_2},$$
$$Cov(\hat{Se}_1, \hat{Se}_2) = \frac{\varepsilon_1}{n_1}, \quad Cov(\hat{Sp}_1, \hat{Sp}_2) = \frac{\varepsilon_0}{n_2}.$$

The equations of the estimators and of the variances-covariances given in the regression, logarithmic, Wald and Fieller intervals are valid substituting $s$ with $n_1$ and $r$ with $n_2$. Regarding the Bootstrap interval, it is necessary to generate $B$ samples with replacement from the case sample and another $B$ samples with replacement from the control sample, and the process is the same as the one described in Section 3.5. Regarding the Bayesian interval, the process is similar substituting $s$ with $n_1$ and $r$ with $n_2$.

The methodology used in this article, both to obtain the *CIs* and to calculate the sample size, can be used to compare other parameters of *BDTs*, e.g. the odds ratios. The odds ratio of a *BDT* is defined as $OR = SeSp/[(1-Se)(1-Sp)]$ and is a measure of the association between the *BDT* and the *GS*. It is easy to check that the ratio of the odds ratios of two *BDTs* is $LR_1^+ LR_2^- / (LR_1^- LR_2^+)$, and therefore from this expression it is possible to deduce *CIs* similar to those given in Section 3 and can also be applied to the same procedure as in Section 5 to determine the sample size necessary to compare the odds ratios of two *BDTs* through a *CI*.

In this manuscript we studied the comparison of the *LRs* of two binary diagnostic tests. When the diagnostic test is quantitative, its accuracy is measured by the area under the *ROC* curve. The *LRs* are related to the equation of the *ROC* curve. Thus, for a single quantitative diagnostic test, for each one of the cut off points $c$ of the estimated *ROC* curve a value for $\hat{S}e$ and a value $1-\hat{S}p$ are obtained, and therefore a value for $\hat{L}R^+$ (and another one for $\hat{L}R^-$). For $\hat{L}R^+$, its numerator $\hat{S}e$ is the "*y*" coordinate of the estimated *ROC* curve, and the denominator $1-\hat{S}p$ is the "*x*" coordinate of the estimated *ROC* curve. The estimator of *LR* for an interval $(c_1, c_2)$ of test values corresponds to the slope of the line segment between $c_1$ and $c_2$ on the estimated *ROC* curve. In the case of two quantitative diagnostic test, for each cut off point of each estimated *ROC* curve, we obtain a value for $\hat{\omega}^+$ and another one for $\hat{\omega}^-$, and therefore it is possible to calculate the *CIs* studied in Section 3.

**Appendix A**

The variances-covariances of all of the parameters were obtained applying the delta method. Let $\boldsymbol{\theta} = (Se_1, Sp_1, Se_2, Sp_2)^T$ be a vector whose components are the sensitivities and the specificities, let $\mathbf{LR} = (LR_1, LR_2)^T$ be a vector whose components are the positive *LRs* or the negative *LRs*, and $\boldsymbol{\omega} = (\omega^+, \omega^-)^T$. The matrix of variances-covariances of $\hat{\boldsymbol{\theta}}$ is

$$\Sigma_{\hat{\boldsymbol{\theta}}} = \left(\frac{\partial \boldsymbol{\psi}}{\partial \boldsymbol{\theta}}\right) \Sigma_{\hat{\boldsymbol{\psi}}} \left(\frac{\partial \boldsymbol{\psi}}{\partial \boldsymbol{\theta}}\right)^T.$$

Regarding the *LRs*, the matrix of variances-covariances of $\hat{\mathbf{LR}}$ is

$$\Sigma_{\hat{\mathbf{LR}}} = \left(\frac{\partial \mathbf{LR}}{\partial \boldsymbol{\theta}}\right) \Sigma_{\hat{\boldsymbol{\theta}}} \left(\frac{\partial \mathbf{LR}}{\partial \boldsymbol{\theta}}\right)^T .$$

Finally, the variance-covariance matrix of $\hat{\boldsymbol{\omega}}$ is

$$\Sigma_{\hat{\boldsymbol{\omega}}} = \left(\frac{\partial \boldsymbol{\omega}}{\partial \boldsymbol{\theta}}\right) \Sigma_{\hat{\boldsymbol{\theta}}} \left(\frac{\partial \boldsymbol{\omega}}{\partial \boldsymbol{\theta}}\right)^T \quad (25)$$

The variance-covariance matrix of $\ln(\boldsymbol{\omega})$ is calculate in a similar way, i.e.

$$\Sigma_{\ln(\hat{\boldsymbol{\omega}})} = \left(\frac{\partial \ln(\boldsymbol{\omega})}{\partial \boldsymbol{\theta}}\right) \Sigma_{\hat{\boldsymbol{\theta}}} \left(\frac{\partial \ln(\boldsymbol{\omega})}{\partial \boldsymbol{\theta}}\right)^T .$$

Performing the algebraic operations in each one of the previous expressions and substituting each parameter with its estimator, we obtain the variances-covariances given in the equations (5), (6), (7) and (10) respectively.

**Appendix B**

The selection of the *CI* with the best asymptotic behaviour was made through the following steps: 1) Choose the *CIs* with the least failures ($CP > 93\%$), 2) Choose the *CIs* which are the most precise (lowest *AL*) and among these those which have a *CP* closest to 95%. The first step in this method establishes that the *CI* does not fail when $CP > 93\%$. The confidence level was set at 95%, i.e. $\gamma = 1 - \alpha = 0.95$ was set as the nominal confidence and, therefore, a nominal error $\alpha = 5\%$. Let $\gamma^*$ be the calculated *CP*, then $\Delta \alpha = \gamma^* - \gamma = \alpha - \alpha^*$, where $\alpha^*$ is the type I error.

Furthermore, the hypothesis test to check the equality of the two LRs is $H_0 : LR_1 = LR_2$ vs $H_1 : LR_1 \neq LR_2$, which is equivalent to checking $H_0 : \omega = 1$ vs $H_0 : \omega \neq 1$. In Step 1, a *CI* fails if $CP \leq 93\%$, i.e. if $\Delta \alpha \leq -2$. In this situation, the type I error of the hypothesis test is $\geq 7\%$, and therefore it is a very liberal hypothesis test and can give false significances. If $\Delta\alpha > 2\%$, i.e. $CP > 97\%$, then the hypothesis test is very conservative (its type I error is very small, $<3\%$), but does not give false significances. Therefore, the choice of the *CI* is linked to the decisions of the hypothesis test, and it is preferable to choose a conservative test rather than a very liberal one (as there will be no false significances due to the fact that its type I error is lower than the nominal one).

**Acknowledgements**

This research was supported by the Spanish Ministry of Economy, Grant Number MTM2016-76938-P.